\documentclass[12pt]{article}

\pdfoutput=1

\usepackage{graphics}
\usepackage{amssymb}
\usepackage{amsmath}
\usepackage{tikz}

\newcommand{\beq}{\begin{equation}}

\newcommand{\eeq}{\end{equation}}
\newcommand{\bea}{\begin{eqnarray}}
\newcommand{\eea}{\end{eqnarray}}
\newcommand{\e}{\mbox{e}}
\newcommand{\R}{{\rm I\!R}}

\begin{document}

\begin{center}
${}$\\
\vspace{60pt}
{ \Large \bf Locally Causal Dynamical Triangulations in \\
\vspace{10pt} Two Dimensions}

\vspace{46pt}

{\sl R. Loll}$\,^{a,b}$
and {\sl B. Ruijl}$\,^{c,d}$

\vspace{24pt}
{\footnotesize

$^a$~Institute for Mathematics, Astrophysics and Particle Physics, Radboud University \\ 
Heyendaalseweg 135, 6525 AJ Nijmegen, The Netherlands.\\ 
{email: r.loll@science.ru.nl}\\

\vspace{10pt}

$^b$~Perimeter Institute for Theoretical Physics,\\
31 Caroline St N, Waterloo, Ontario N2L 2Y5, Canada.\\
{email: rloll@perimeterinstitute.ca}

\vspace{10pt}

$^c$~Leiden Institute of Advanced Computer Science, Leiden University\\ 
Niels Bohrweg 1, 2333 CA Leiden, The Netherlands.\\
{email: b.j.g.ruijl@liacs.leidenuniv.nl}\\

\vspace{10pt}

$^d$~Theory Group, NIKHEF\\ 
Science Park 105, 1098 XG Amsterdam, The Netherlands.\\
{email: benrl@nikhef.nl}\\

}
\vspace{48pt}

\end{center}

%\addtolength{\baselineskip}{0.20\baselineskip}
%\vspace{2cm}

\begin{center}
{\bf Abstract}
\end{center}

\noindent We analyze the universal properties of a new two-dimensional quantum gravity model
defined in terms of Locally Causal Dynamical Triangulations (LCDT). Measuring the Hausdorff
and spectral dimensions of the dynamical geometrical ensemble, we find numerical evidence that
the continuum limit of the model lies in a new universality class of two-dimensional quantum gravity
theories, inequivalent to both Euclidean and Causal Dynamical Triangulations.
\vspace{12pt}
\noindent

%\vfill

\newpage

\section{Causal quantum geometry in two dimensions}
\label{intro:sec}

The approach of Causal Dynamical Triangulations (CDT) \cite{physrep} provides concrete evidence that 
one must include causal, Lorentzian properties in the nonperturbative gravitational path integral
in order for the associated quantum gravity theory to possess a classical limit. 
This should be contrasted with purely Euclidean constructions where the path integral
is taken over geometric configurations which represent four-dimensional ``spacetimes", but 
do not contain any information about time, light cones or causality. The problem with the latter appears to be
that -- quite independent of the elementary ``building blocks" used for constructing individual Euclidean 
configurations -- their nonperturbative sum is completely dominated by highly degenerate
objects whose superposition never leads to a four-dimensional extended universe on macroscopic
scales, no matter how one looks at it.

The Euclidean precursor of CDT, so-called Dynamical Triangulations (DT), is a case in point. 
In both DT and CDT quantum gravity one looks for scaling limits of regularized path integral 
expressions, where curved geometries are represented by triangulated, piecewise flat manifolds.
However, the infinite-volume limit of the Euclidean theory is dominated by degenerate phases with no obvious
physical interpretation in terms of general relativity \cite{edt}, 
and phase transitions are of first order \cite{firstorder1,firstorder2}.
By contrast, one of the phases of CDT quantum gravity is characterized by a quantum geometry
which on large scales exhibits properties of a four-dimensional de Sitter space \cite{desitter}, and 
the phase space of CDT contains a whole line of second-order critical points \cite{secondorder}, 
which are being investigated as natural candidates for defining the searched-for continuum theory \cite{cdtrg}.

At the inception of the CDT approach, the toy model version of the theory in two spacetime dimensions
played an important role: the nonperturbative CDT path integral over geometries can in this case be evaluated
analytically \cite{al}, and the resulting two-dimensional quantum gravity theory compared with
the much-studied theory of two-dimensional Euclidean quantum gravity, which likewise can be formulated
and solved exactly in terms of dynamical triangulations (see \cite{edtreviews} for reviews). 
The two theories turn out to be inequivalent, and are characterized by different critical exponents (see \cite{alet}
for a comparison). In terms of continuum formulations, they lie in the universality class of Liouville quantum
gravity \cite{liouville} for DT and two-dimensional, projectable Ho\v rava-Lifshitz gravity \cite{hl2} for CDT. 
These models provide
the first completely explicit example that ``signature matters" in the context of the nonperturbative
gravitational path integral, aka the ``sum over histories". As mentioned above, we now have good evidence
that the same is true for the physically relevant case of quantum gravity in four dimensions. 

This paper will expand on the theme of two-dimensional quantum gravity as an interesting testing ground for
quantum gravity proper, where both analytical and numerical solution methods can be employed and compared.
We will study the two-dimensional implementation of a recently introduced version of CDT, which goes
by the name of ``Locally Causal Dynamical Triangulations (LCDT)" \cite{jordanthesis} or 
``CDT without preferred foliation" \cite{jordanloll}. The path integral in these CDT models is performed over a
class of piecewise flat Lorentzian geometries that is enlarged compared to standard CDT quantum
gravity. The geometries are still causal, in the sense of having a well-defined light cone structure 
everywhere, but are not required to have the preferred (discrete) proper-time slicing 
characteristic of standard CDT configurations. An in-depth numerical investigation of locally causal DT in
three spacetime dimensions found nontrivial evidence that key results of CDT quantum gravity, 
including the volume distribution of three-dimensional de Sitter space are reproduced in this
generalized causal theory \cite{jordanloll}. 

This is an important and concrete piece of evidence that for a judicious choice of the bare coupling constants of the theory,  
LCDT and CDT quantum gravity lead to equivalent continuum theories. We would like to
investigate whether the same is true in two spacetime dimensions. Although this toy model is arguably even less
representative of full gravity than the three-dimensional model, the properties of ``quantum geometry" are
much simpler to analyze in dimension two and may give us a hint of why the two causal theories are
equivalent or not, as the case may be. 

Since in terms of its configuration space the locally causal model
lies in between DT and CDT quantum gravity, solving it will give us a better understanding of the universality
classes of theories of quantum geometry in two dimensions. The CDT universality class has so far proven
to be quite robust: inclusion of a higher-curvature term \cite{fgk1}, a decoration by arches along spatial
links (tantamount to including a restricted class of ``baby universes") \cite{fgk2,ambips}, 
an explicit inclusion of a finite number of baby universes within a string field-theoretic setting
based on CDT \cite{sft-cdt} (see also \cite{ambjornbudd}), or starting from a
conceptually rather different Hamiltonian ``string bit model" \cite{durhuuslee} all lead to the same
scaling limit. In the absence, to date, of an analytic solution of locally causal DT in two dimensions, we
will present below the results of a numerical investigation. We have examined several observables and 
measured two critical exponents, 
the expectation values of the Hausdorff and the spectral dimension of quantum spacetime, to try to understand
whether they coincide with those of DT or CDT quantum gravity, or perhaps signal yet another, new universality
class of two-dimensional quantum geometry. 
  
We begin our analysis by introducing the locally causal DT model and its geometry in Secs.\ \ref{lcdt:sec} and
\ref{invest:sec}, and outline the set-up for the Monte Carlo simulations in Sec.\ \ref{setup:sec}. Closed
timelike curves and their role in LCDT are described in Sec.\ \ref{ctc-sec}. Sec.\ \ref{obs:sec} deals with observables,
including the volume profile, a characterization of the quantum geometry in terms of minimal loops, 
and our results for the Hausdorff and spectral dimensions. Our conclusions are presented in Sec.\ \ref{concl:sec}.
The Appendix contains some details on the geometric shape of typical LCDT configurations.

\section{Locally causal dynamical triangulations}
\label{lcdt:sec}

We will first derive an expression for the gravitational action in locally causal DT in two dimensions. 
Our starting point is the two-dimensional Einstein-Hilbert action in the continuum,
\beq
\label{contact}
S=\kappa \int d^2x \sqrt{|g|} (R-2\Lambda),
\eeq
where $\kappa$ is the inverse of Newton's constant, $\Lambda$ the cosmological constant,
and $g$ denotes the determinant of the Lorentzian spacetime metric $g_{\mu\nu}$. The integral over
the scalar curvature $R$ is topological and will not play a role in the path integral construction,
since the spacetime topology will be held fixed. Dropping the $R$-term leaves us with just the volume term. 
Absorbing the (dimensionless) gravitational constant into the cosmological constant $\Lambda$, the path
integral and its regularized counterpart in terms of dynamical triangulations read
\beq
\label{pathints}
\int{\cal D}[g]\, \e^{-i\Lambda\int\! d^2x \sqrt{|g|}}\; \longrightarrow\; \sum_T \frac{1}{C(T)}\,\e^{-i\lambda V_2(T)},
\eeq
where the (formal) integration over diffeomorphism equivalence classes $[g]$ of metrics on the left-hand
side has been replaced by a sum over inequivalent triangulations $T$ with the usual DT measure
involving the order $C(T)$ of the automorphism group of $T$. The constant $\lambda$ on the right-hand
side is the bare cosmological coupling of the regularized theory, and $V_2(T)$ denotes the spacetime volume 
of the triangulation $T$. 

\begin{figure}[ht]
\centering
\scalebox{0.5}{\includegraphics{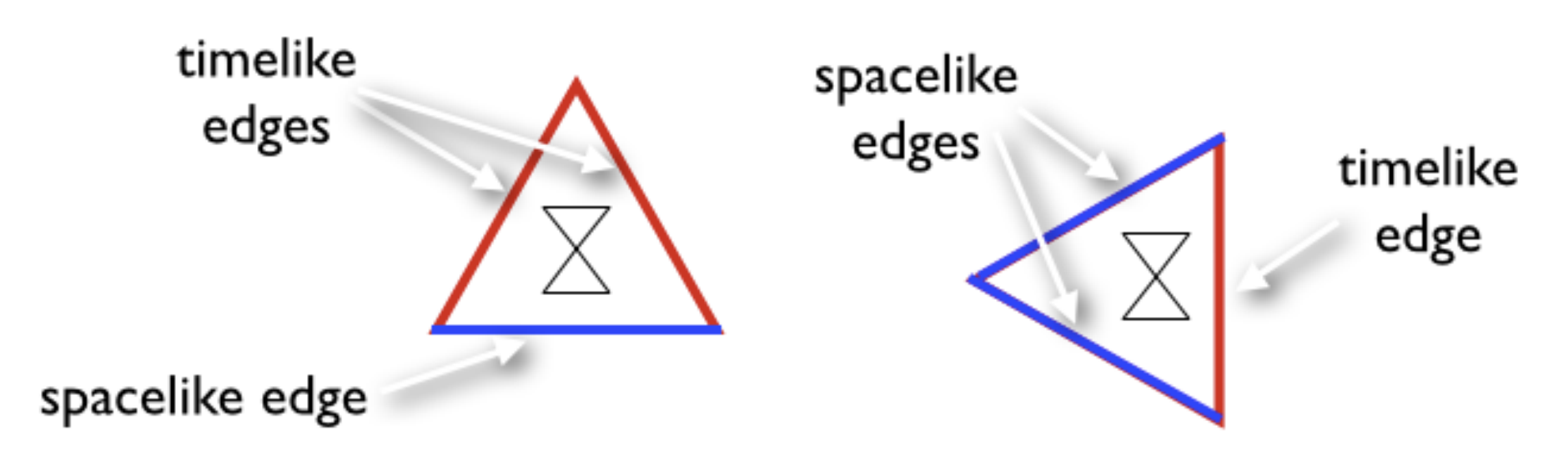}}
\caption[phase]{The two elementary building blocks of (1+1)-dimensional LCDT, with light cones 
indicated: $stt$-triangle $\Delta_{stt}$ with one space- and two timelike edges (left) and $sst$-triangle $\Delta_{sst}$
with one time- and two spacelike edges (right). The colour-coding for spacelike edges is blue, and for timelike ones red.
}     
\label{twotri}
\end{figure}

As described in \cite{jordanloll}, LCDT in 1+1 dimensions uses two types of triangular Minkowskian
building blocks, to allow for the construction of geometries without a preferred time foliation (see Fig.\ \ref{twotri}).
The usual CDT simplex $\Delta_{stt}$ with one spacelike and two timelike edges is supplemented by another
two-simplex $\Delta_{sst}$ with one timelike and two spacelike edges. The squared edge length of all
spacelike links is $l_s^2\! =\! a^2$ and that of timelike links $l_t^2\! =\! -\alpha a^2$, in terms
of the lattice cutoff $a$ and the ratio $\alpha >0$ of the two quantities. To determine the 
spacetime volume $V_2(T)$
of a triangulation $T$ assembled from these building blocks, we simply need to count their numbers
$N_{stt}(T)$ and $N_{sst}(T)$ and compute the volumes of both $\Delta_{stt}$ and $\Delta_{sst}$.
The latter are determined in a straightforward way 
from the values of the edge lengths of the two triangles, and are given by
\beq
\text{vol}(\Delta_{stt})= \frac{\sqrt{4\alpha +1}}{4}\, a^2  ,\;\;\;\; \text{vol}(\Delta_{sst}) =\frac{\sqrt{\alpha (\alpha+4)}}{4}\, a^2.
\eeq 
To be able to perform Monte Carlo simulations of the system, we analytically continue the parameter $\alpha$
to $-\alpha$ in the lower-half complex plane according to the usual CDT prescription \cite{3d4d}. The resulting
real expression for the Wick-rotated regularized path integral in two dimensions is
\beq
\label{pilcdt}
Z(\lambda)= \sum_T \frac{1}{C(T)}\,\e^{-\lambda a^2(N_{stt} \frac{\sqrt{4\alpha -1}}{4}+N_{sst} 
\frac{\sqrt{\alpha (4-\alpha)}}{4})} .
\eeq
Note that for both triangle volumes to be positive, $\alpha$ must satisfy the inequality 
$1/4 \! <\! \alpha\! < \! 4$. The limiting value $\alpha\! =\! 1/4$ corresponds geometrically to a collapse 
of the $stt$-triangles to zero volume, whereas for $\alpha\! =\! 4$ the $sst$-triangles are collapsed.
In the isotropic case, $\alpha\! =\! 1$, both triangles after Wick rotation are equilateral and identical. 
All LCDT simulations presented below were performed for $\alpha\! =\! 1$. 
\begin{figure}[t]
\centering
\scalebox{0.5}{\includegraphics{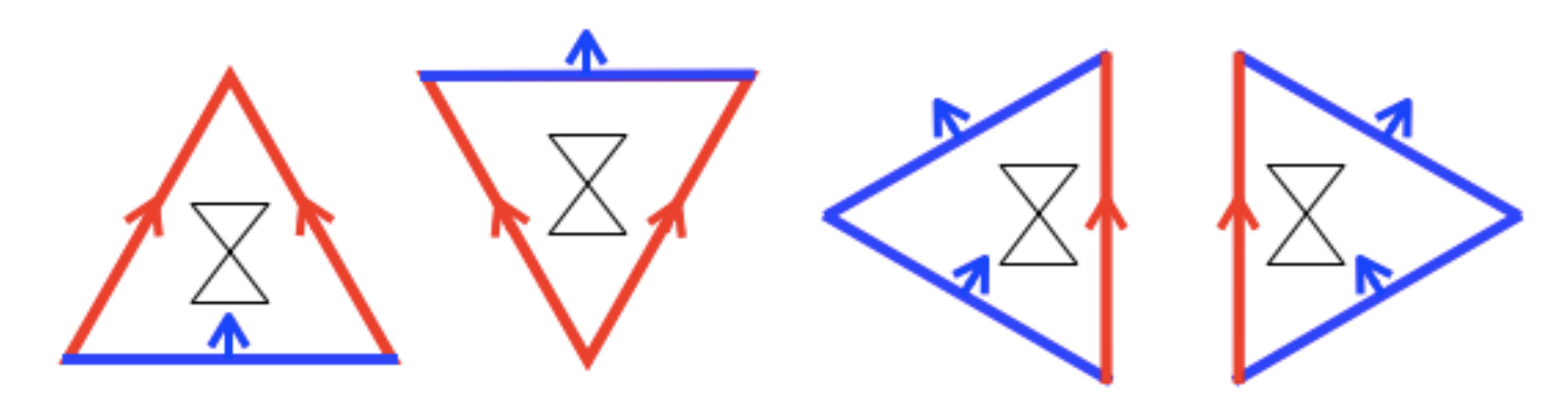}}
\caption[phase]{Two time orientations are possible for each triangle type, as indicated by the future-pointing
arrow assignments.
}     
\label{triorient}
\end{figure}

\begin{figure}[t]
\centering
\scalebox{0.5}{\includegraphics{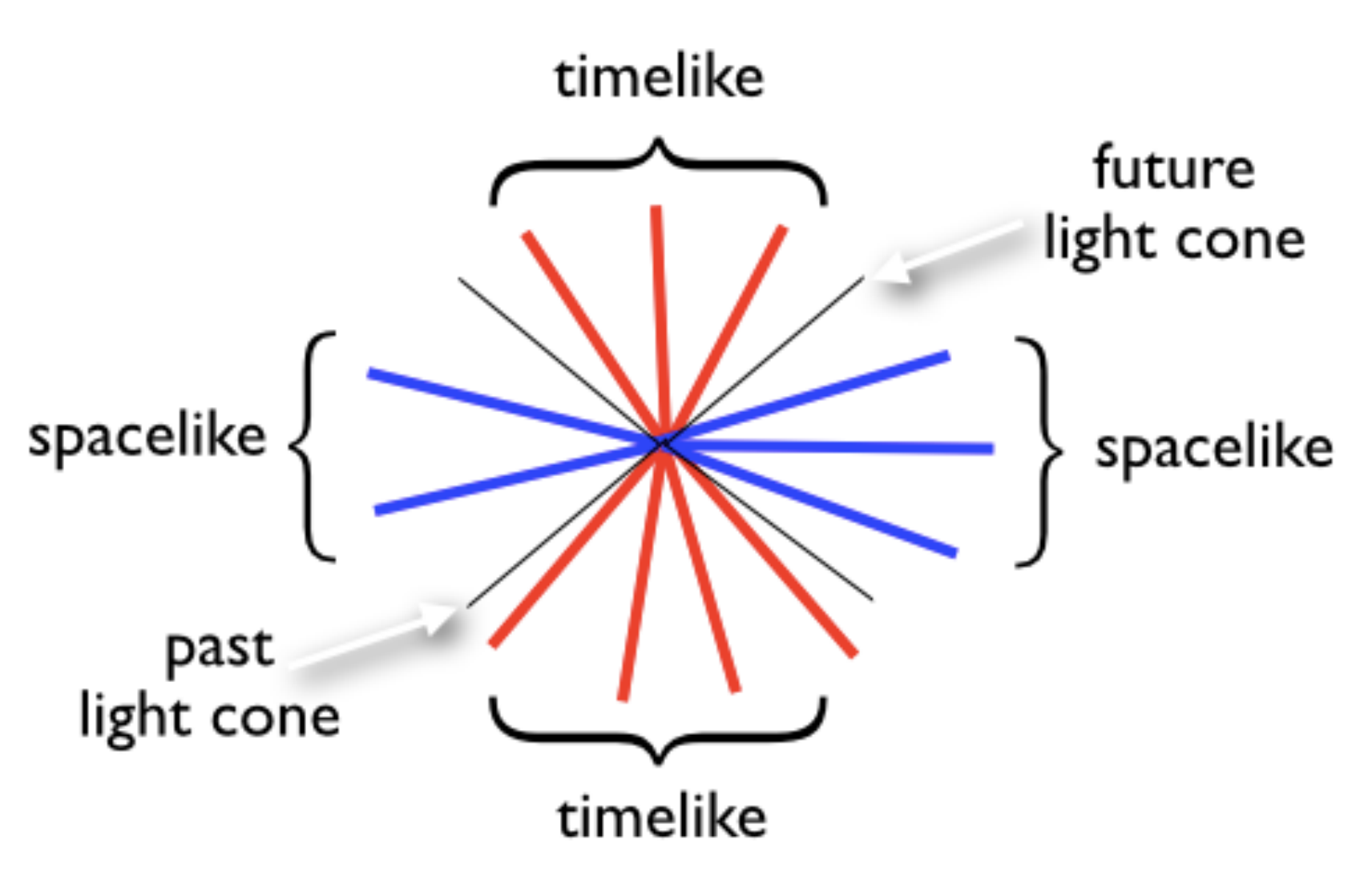}}
\caption[phase]{Local causality implies the existence of one future and one past light cone at each point.
The edges meeting at a vertex are always arranged in four groups 
of alternating type, time- or spacelike (with at least one edge in each group), 
as depicted. In the figure, ``time" is pointing upwards, and 
the thin lines indicate the light cones located at the central vertex.
}     
\label{vertexconf}
\end{figure}

Building blocks of the two triangle types are assembled into simplicial manifolds $T$ by ``gluing" them together
pairwise along boundary edges, where a timelike edge can only be glued to another timelike edge,
and a spacelike edge only to another spacelike edge. Note that with respect to some overall flow of time, each of the
two building blocks can appear with two different time orientations, which can be indicated by
arrow assignments as illustrated in Fig.\ \ref{triorient}. By definition, all arrows are future-pointing. 
Note that once a single triangle in a triangulation has been given a specific time orientation, the orientation of its
neighbours, and of its neighbours' neighbours, etc. is also fixed, because the arrows on shared edges have to match.

Local causality is incorporated in the gluing rules by stipulating 
that before the analytic continuation there should be exactly one future and one past light cone at each
interior point of the triangulation \cite{jordanloll}. This condition is always satisfied at an interior
point of a triangle, because up to diffeomorphisms the metric by construction is given by the Minkowski metric. 
It is also satisfied at points along edges where two triangles meet, unless the point happens to be a vertex, as
can be seen by inspecting the geometry of the building blocks in Fig.\ \ref{triorient}. The
requirement is only nontrivial at the vertices of the triangulation. When expressed in terms of the
edges meeting at a vertex, it implies that they should come in four groups of alternating type (time- or spacelike)
when going around the vertex once (Fig.\ \ref{vertexconf}), which imposes corresponding restrictions
on the triangles meeting at the vertex. 

\begin{figure}[t]
\centering
\scalebox{0.4}{\includegraphics{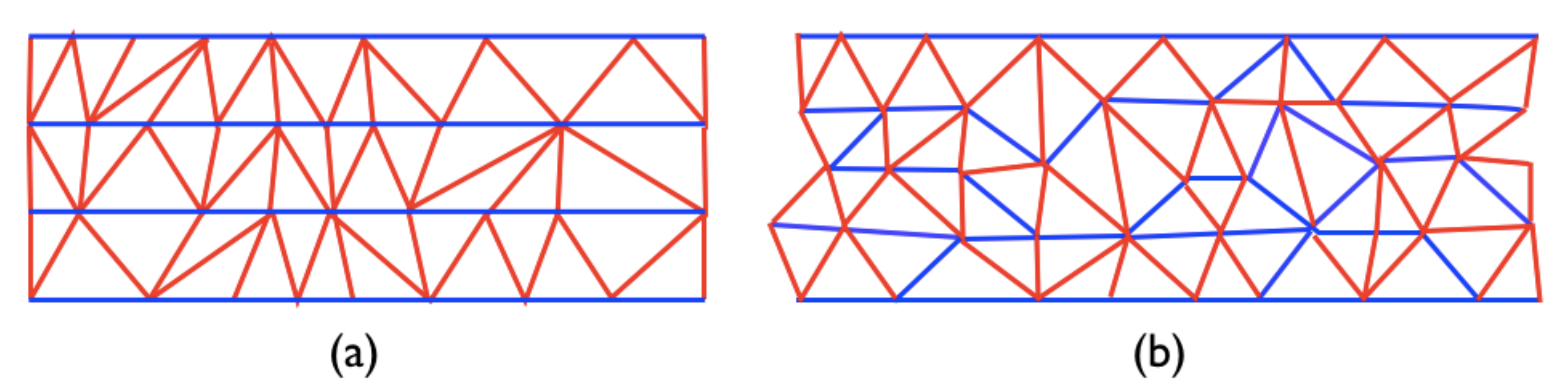}}
\caption[phase]{Two strips of dynamically triangulated spacetimes with an initial (bottom) and a final
spatial boundary (top). (a) In standard CDT only $stt$-triangles are used. There are two spatial
edges meeting at each internal vertex, giving rise to the characteristic preferred proper-time slicing
in terms of consecutive lines of spatial links. (b) LCDT works with both $stt$- and $sst$-triangles,
allowing for a more general vertex structure, without preferred slicing.
}     
\label{DTstrips}
\end{figure}

The generic vertex structure in ordinary CDT -- which uses only building blocks of type
$\Delta_{stt}$ -- is also of this type, but by construction there is only
a single spacelike edge each on the left and the right of the light cone, and the pair of these
spacelike links
forms part of a preferred slice of constant integer time. It is precisely the generalized vertex structure 
in locally causal DT that allows for configurations without this preferred time slicing. Fig.\ \ref{DTstrips}
illustrates the difference between a piece of causal DT and one of locally causal DT.

\section{Properties of LCDT in two dimensions}
\label{invest:sec}

Our task will be to investigate the properties of the path integral (\ref{pilcdt}) in the continuum limit, where
the sum is taken over an ensemble of triangulations of fixed topology, obeying local simplicial manifold 
conditions\footnote{in dimension 2: 
each internal edge is shared by exactly two triangles, and any two triangles share at most one edge}
and with the vertex structure of locally causal DT described above. 
At this stage, there is no known exact solution of the continuum dynamics of LCDT quantum gravity in two 
dimensions; one difficulty is precisely the absence of a distinguished notion of time in terms of the lattice structure itself, 
which prevents the straightforward introduction of a transfer matrix used 
previously in solving CDT \cite{al,fgk1,fgk2,ambips}. 

Since the configurations have two
different kinds of edges, they can be thought of as a particular kind of two-coloured graphs, whose
properties one may try to understand in a systematic way in the sense of enumerative combinatorics.
The subgraph consisting of spacelike links only has the form of a stack of ``bubbles", which on their inside are
decorated with timelike links (see \cite{hoekzema,jordanloll} for definitions and discussions of this substructure). 
The LCDT model may be supplemented by additional conditions, which restrict the
type of (self-)overlaps among these bubbles that are allowed.  
Since the bubbles are extended structures, these additional rules have a nonlocal character. 
They are motivated by the finding that for spatially compact boundary conditions in two
dimensions the local condition of vertex causality described above does not 
imply global causality in the sense of the absence of (a specific class of) closed timelike curves
(see Sec.\ \ref{ctc-sec} for an explicit example and further discussion).
In turn, the presence of such curves is related to the presence of overlapping bubbles.

In our simulations both space and time will
be compact. The topology of space will be a circle and time will be cyclically identified, which means that
spacetime is topologically a torus $T^2$. Since we are primarily interested in ``bulk" properties of the
geometry, this choice is technically convenient:
all vertices are interior vertices, on which vertex causality will
be imposed, and the action is the one appearing in the path integral expression (\ref{pilcdt}), 
without the need for adding any boundary terms. 

The functional form of the Euclidean action, schematically
given by $S\! =\! c_1 N_{stt} +c_2 N_{sst}$, for two positive constants $c_1$, $c_2$, is the most
general one linear in ``counting variables". These are the variables counting simplices of a
particular dimension and type in a given triangulation $T$: the numbers $N_0(T)$ of vertices, 
$N_s(T)$ of spacelike edges and $N_t(T)$ of timelike edges, as well as the numbers
$N_{sst}(T)$ and $N_{stt}(T)$ already introduced earlier. Our statement follows from the fact
that these five variables are subject to three constraints, 
\begin{equation}
N_0-N_s-N_t+N_{sst}+N_{stt}\! =0,\;\;
N_{sst}+2 N_{stt}-2 N_t\! =0,\;\;
N_{stt}+2 N_{sst}-2N_t\! =0,
\end{equation}
which must be satisfied on each configuration $T$ with torus topology.

We finally note that, at least in the absence of additional
constraints on the bubble configurations, toroidal boundary conditions
introduce a duality into the two-dimensional LCDT system. The duality transformation 
consists in swapping 
simultaneously the assignments ``timelike" and ``spacelike" of all edges in a given triangulation,
which will convert all $stt$-triangles into $sst$-triangles and vice versa.
An admissible triangulation (one that satisfies the local conditions of a simplicial manifold and
vertex causality) will under this transformation be mapped to another admissible triangulation with the same
topology, with the roles of time and
space interchanged. Of course, for $\alpha\! \not=\! 1$ the Boltzmann weights of a triangulation and its
dual will in general be different.

\section{Numerical set-up}
\label{setup:sec}

We have used Monte Carlo techniques to sample the partition function or Euclideanized path integral
(\ref{pilcdt}) of locally causal dynamical triangulations, and compute expectation values of selected observables.
An important ingredient are a set of Monte Carlo moves, which take the form of local changes in the simplicial 
geometry, and are designed to get us around the configuration space of the model by way of a Markov process.
The four types of move we have worked with will be described below. 
Further technical details on their
implementation may be found in \cite{ruijl}. Additional references on Monte Carlo moves in the context of
causal dynamical triangulations are \cite{jordanthesis,physrep}. 

\begin{figure}[t]
\centering
\scalebox{0.43}{\includegraphics{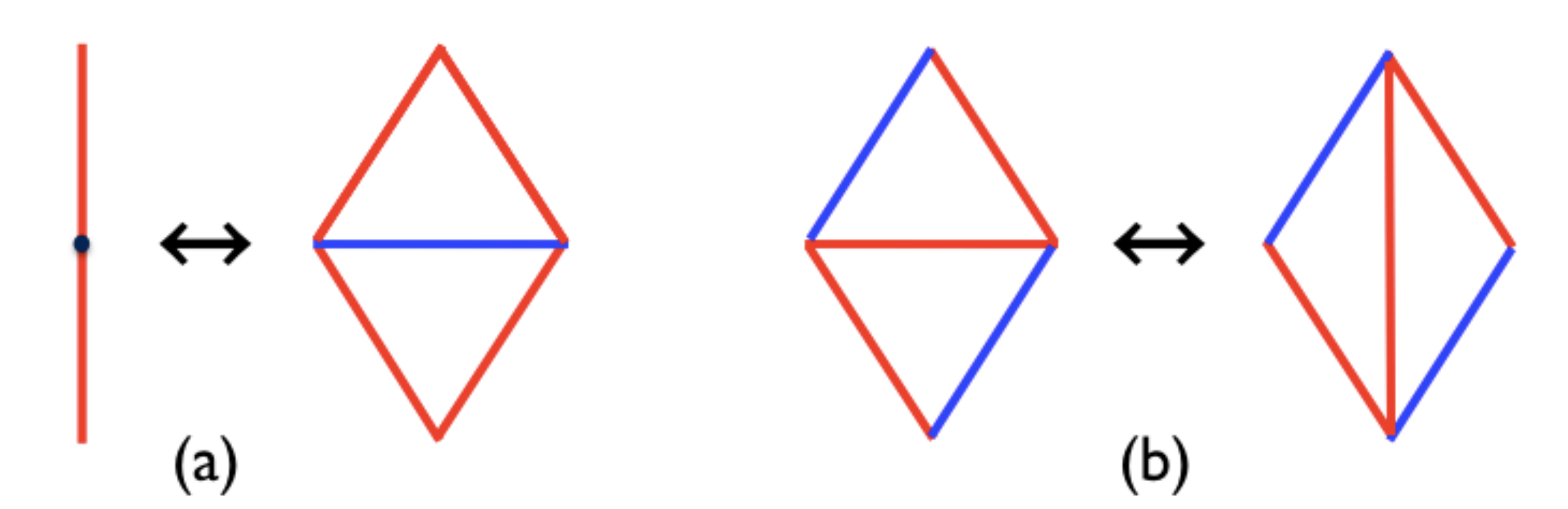}}
\caption[phase]{Two types of Monte Carlo moves used in locally causal DT: (a) example of a (0,2)-move and its inverse, also
called a ``link collapse"; (b) example of a (2,2)- or flip move and its inverse.
}     
\label{collapseflip}
\end{figure}

The first type of move is a generalization of the (0,2)-move used in CDT, in which two adjacent $stt$-triangles that
share a spacelike link are
created simultaneously. The local starting configuration consists of a pair of timelike links, belonging to 
opposite sectors of a vertex (one link from inside the past and the other from inside the future light cone), see 
Fig.\ \ref{collapseflip}(a). This move is compatible with the time slicing of CDT geometries. Since this compatibility is no
longer a requirement in locally causal DT, we will also use the colour-reversed counterpart of this move, where
two adjacent $sst$-triangles that share a timelike link are created from a pair of spacelike links, 
this time from opposite spacelike sectors of the light cone at the central vertex.

Also the second type of move, the (2,2)- or ``flip" move generalizes a local move already employed in CDT. It consists in
flipping the diagonal inside a rhombus made of a pair of adjacent triangles. The version depicted in 
Fig.\ \ref{collapseflip}(b) is the one also permitted in CDT. In our simulations of locally causal DT, we will in addition use
the move with the opposite assignments of time- and spacelike edges. These are the only two flip moves compatible 
with vertex causality, if the character of the flipped edge remains unchanged. Two more flip moves are possible
if the flipped diagonal link is allowed to change from time- to spacelike or vice versa.

\begin{figure}[t]
\centering
\scalebox{0.43}{\includegraphics{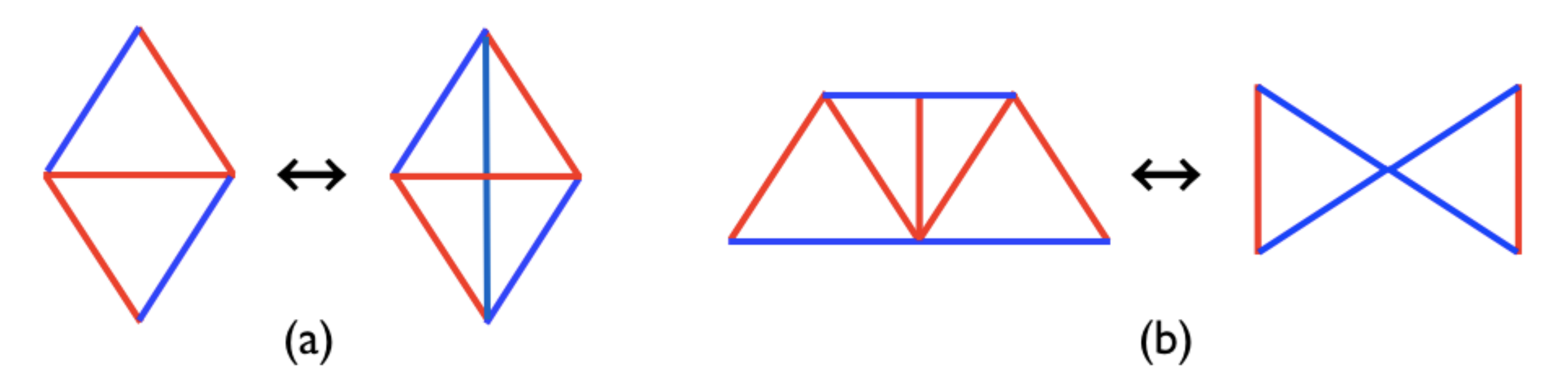}}
\caption[phase]{The two remaining types of Monte Carlo moves used in locally causal DT: (a) example of a (2,4)-move and its inverse; (b) the pinching move and its inverse.
}     
\label{pinchother}
\end{figure}

Another type of move we have used in the simulations is a (2,4)-move, where the starting point is again given
by a pair of adjacent triangles. A new configuration with identical boundary is obtained by ``subdividing" the rhombus
with another diagonal, thereby creating a four-valent vertex at the centre, see Fig.\ \ref{pinchother}(a) for an example. 
In order for vertex causality to be satisfied at the new vertex, the added diagonal has to be of opposite (time-/spacelike) 
type to the one already present. 
Eight variations of the (2,4)-move (and its inverse) are possible, depending on the type and orientation of the
initial triangle pair, but six of them are equivalent to performing a (0,2)-move, which was already discussed above.

Lastly, we employ a ``pinching" move, which is entirely new and not present in standard CDT, and was previously
described in \cite{jordanthesis}. The initial local
configuration for this move looks like a piece of regular CDT configuration, consisting of four $stt$-triangles
forming a strip bounded above and below by line segments made out of spacelike links. The move pinches
those two segments together in a single point, resulting in a pair of $sst$-triangles, as illustrated by 
Fig.\ \ref{pinchother}(b). In the simulations we also use the colour-reversed version of this move.

In all cases, it is understood that whenever one of these moves is proposed by the computer algorithm, 
it will always be rejected if it violates either vertex causality or the simplicial manifold condition.  
The (overcomplete) set of 
these moves is likely to be ergodic, but we do not have a formal proof at this stage.
The explicit proof may depend in subtle ways on the details of the
ensemble, in particular, on excluding classes of bubble configurations associated with 
specific closed timelike curves that lead to unwanted global acausal behaviour.

We have run several kinds of cross check on the Monte Carlo simulations: firstly, that the acceptance
rates of moves and their inverses are approximately the same, and secondly, that the frequency of occurrence
for configurations with very small volume is compatible with the frequency predicted by the Boltzmann
distribution. Lastly, our set-up has an easy way to implement a CDT limit, which we can use to
cross-check measured CDT results for the dynamical dimensions with the theoretical results available.

Note that apart from the flip move, all Monte Carlo moves described above alter the number of triangles in
the triangulation, and therefore its two-volume. Since it is convenient from the numerical point of view to keep
the total volume fixed, at least approximately, we use the standard DT prescription where the volume $V_2$
is allowed to vary in a narrow interval around a fixed target volume $V_2^{(0)}$. This is achieved by adding 
a quadratic term $\delta\, (V_2^{(0)}\! -\! V_2)^2$ to the action, with a parameter $\delta >0$ determining
the width of the interval.\footnote{Note that for our standard choice $\alpha\! =\! 1$, $V_2$ is proportional to the
number $N_2$ of triangles.} We tune the cosmological
constant $\lambda$ such that the measured volumes are distributed symmetrically around $V_2^{(0)}$,
and only collect data from triangulations that have precisely this target volume. The simulations are run
such that there is about one sweep of length $L\approx 10^6$ between successive measurements.

As mentioned before, there is a straightforward way to obtain ordinary CDT simulations in our set-up. It
consists in setting $\alpha\! =\! 1/4$ in the action (\ref{pilcdt}), while maintaining the constraint
$N_2\!\equiv\! N_{sst}+N_{stt}\! =\! constant$. Since the term proportional to $N_{stt}$ in the action now
vanishes, $stt$-triangles can be created ``at no cost" during the Monte Carlo simulation, while the number
of $sst$-triangles will diminish accordingly, thereby lowering the value of the Euclidean action. As a result,
the triangulations quickly become pure CDT configurations, consisting only of building blocks $\Delta_{stt}$. 
Below in Sec.\ \ref{obs:sec} we measure the Hausdorff and spectral dimensions 
of CDT quantum gravity. In addition to setting $\alpha\! =\! 1/4$, we will disable the $sst$-triangles completely,
to make sure that any fluctuations with nonvanishing $N_{sst}$ are eliminated.

\section{Closed timelike curves}
\label{ctc-sec}

For most of our measurements, the ensemble of locally causal triangulated geometries on which the dynamics takes
place will consist of spacetimes of torus topology satisfying simplicial manifold conditions 
and vertex causality. However, for some purposes,
when considering the time evolution of observables, it is convenient to restrict this ensemble further,
because of the appearance of a particular class of closed timelike curves. 
\begin{figure}[t]
\centering
\scalebox{0.55}{\includegraphics{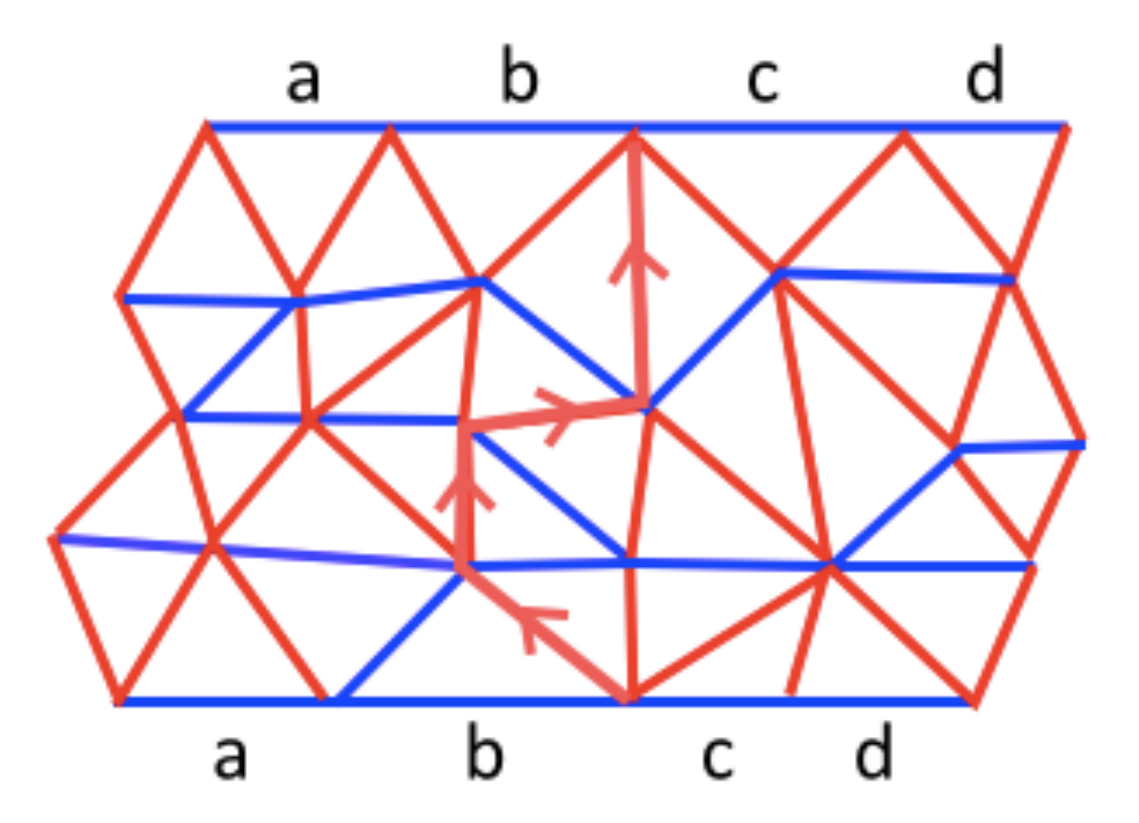}}
\caption[phase]{A piece of LCDT, with an initial and a final spatial boundary (blue edges, bottom and top). 
When the two spatial boundaries are identified as indicated by the letters, the future-oriented timelike curve
running from bottom to top (timelike edges with arrows) becomes a timelike cycle, as defined in the text.
}     
\label{tcurve}
\end{figure}

To explain this issue further, we introduce the notion of a {\it timelike cycle} in a locally causal
triangulation. By this we shall mean a contiguous set of timelike links which together form a {\it non-contractible}
loop of topology $S^1$, without self-crossings or self-overlaps (Fig.\ \ref{tcurve}). In addition, whenever the loop
passes through a vertex, the two timelike links of the loop meeting at the vertex must lie in opposite light cones, 
never inside the same (half of the) light cone. A {\it spacelike cycle} is defined analogously in terms of
spacelike edges, and it is also required to be non-contractible. 
In the spacelike case, the cycle has to cross at each vertex from one spacelike sector outside the light 
cone to the opposite one.

In standard CDT in 1+1 dimensions with its preferred time slicing (c.f.\ Fig.\ \ref{DTstrips}a), 
spacelike cycles only exist when 
space is compactified to a circle. In this case they simply coincide with
the one-dimensional spatial slices at integer proper time. Likewise, timelike cycles only exist in CDT when 
time is compactified, in which case they can be thought of as a particular\footnote{They are particular
in the sense that one could also consider paths that run not only along the
edges of a triangulation, but also through the interiors of triangles.} lattice realization of 
closed timelike curves. As a consequence of
how they traverse the light cones at vertices, they are also time-oriented, either in positive or negative time direction.
As usual, the reason for compactifying time in CDT simulations is merely one of convenience, and the appearance of
closed timelike curves is an inevitable side effect, which is not expected to have much influence
on the measurement of ``bulk" observables like the dynamical dimensions considered below.\footnote{Obviously, 
in any concrete implementation the
dependence of observables on boundary conditions and other finite-size effects should always be monitored.
Although CTCs are unphysical classically, their inclusion in the regularized path integral does not a priori imply 
unphysical behaviour of the final continuum theory.}
\begin{figure}[t]
\centering
\scalebox{0.5}{\includegraphics{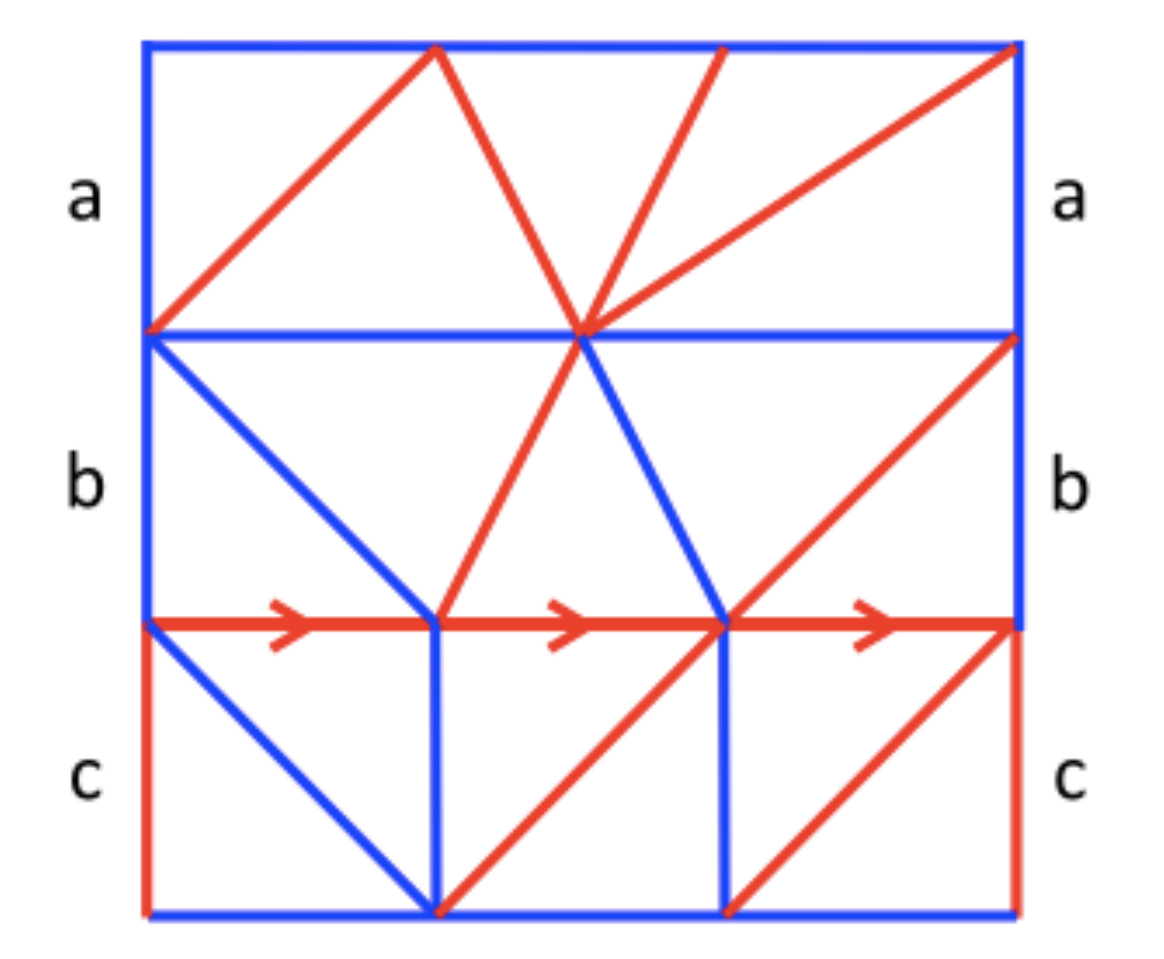}}
\caption[phase]{A piece of locally causal DT, with an initial and a final spatial boundary (blue edges, bottom and top). 
Compactifying space as indicated by the letters results in a closed, future-oriented timelike curve (timelike edges
with arrows), which intersects neither the initial nor the final boundary.
}     
\label{ctc}
\end{figure}

The closed timelike curves we are primarily concerned about in LCDT are not the ones winding around
the compactified time direction, but around the spatial direction, and which would still be
present if spacetime was a cylinder $I\times S^1$ (with compact spatial slices) instead of a torus, see Fig.\ \ref{ctc}
for an example. In the remainder of this work, when we talk about closed timelike curves (CTCs), we will mean
only this class of timelike cycles. 
It turns out that local vertex causality does not preclude the presence of these curves, although it does
prevent the occurrence of {\it contractible} timelike loops \cite{hoekzema}. One way of finding
them is by running an algorithm that assigns time labels to vertices of a given locally causal DT. 
Of course, an explicit choice of time has to be made in the LCDT model, because -- unlike in
usual CDT -- there is no preferred lattice substructure one can refer to as a natural time label. 

The prescription for assigning time labels to vertices in a given geometry
we have used in the present work is to pick a spacelike cycle in the geometry and define it to be ``space
at time $t\! =\! 0$". Vertices lying in the future of this slice are then successively assigned time labels, which are
computed as the average distance of the vertex $v$ to the initial slice along any oriented timelike path from
the slice to $v$ (see \cite{ruijl} for more details on the algorithm). 
The distance along any given path is simply given by the number of timelike links it contains, the
so-called {\it link distance}.
Note that the time label of a vertex will in general not be an integer. Once the vertices are labelled, one
can in a straightforward way also associate time labels with edges or more extended regions like spacelike cycles
by averaging over the time labels of the vertices contained in them. In standard CDT, this prescription
reproduces the usual integer proper-time slicing.

The algorithm implementing the vertex labelling breaks down when it encounters a CTC, like
that depicted in Fig.\ \ref{ctc}, because for any vertex lying on the curve or in its future there will be infinitely 
many timelike paths connecting it to the initial slice.
We conclude that in LCDT, at least when space is compactified, local causality does not imply global causality in the sense of
an absence of CTCs.

\begin{figure}[t]
\centering
\scalebox{0.55}{\includegraphics{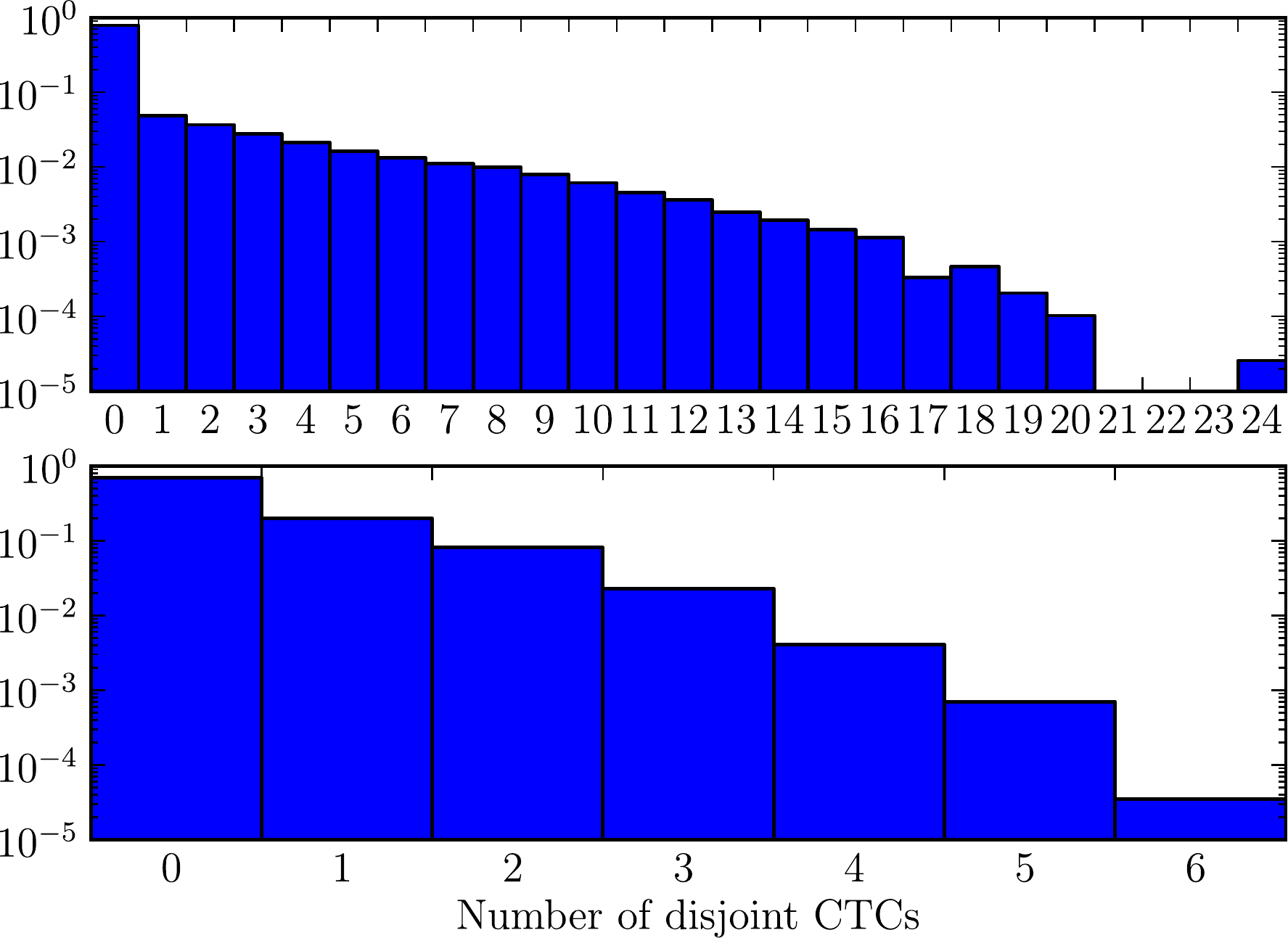}}
\caption{The frequency of disjoint timelike loops in 48.000 sweeps. 
The upper histogram is for $N_2\! =\! 10.000$ and the lower one for $N_2\! =\! 100.000$.
}     
\label{ctc_freq}
\end{figure}

Whether or not the presence of closed timelike curves has any consequences for the 
continuum limit of the model is a priori unclear. We have performed a number of measurements
to get a better quantitative idea of how many CTCs there are, depending on the size of the
triangulation. 
Fig.\ \ref{ctc_freq} shows two histograms of the frequencies of disjoint CTCs (CTCs without mutual 
overlap\footnote{These curves form a subclass of CTCs; the number of all CTCs can be considerably larger,
especially when there are many disjoint CTCs.}). 
We observe that the typical number of disjoint closed timelike curves present in a given configuration goes down
significantly when the volume is increased tenfold from $N_2\! =\! 10.000$ to $N_2\! =\! 100.000$ triangles.
On the other hand, the ratio
of triangulations which contain any CTCs at all increases from about 21\% to 30\%. 
We therefore have no indication that CTCs disappear completely as the volume grows.
One contributing factor is presumably that it takes much longer to break up a CTC by a local Monte Carlo move 
when the CTCs become highly diluted in a triangulation.

\section{Observables}
\label{obs:sec}

Having introduced both the theoretical and numerical set-up of locally causal dynamical triangulations
in 1+1 dimensions, we will now discuss the measurements of several observables in this model of
quantum geometry, concentrating on the isotropic case $\alpha\! =\! 1$. 
Our main aim is to understand whether the model's continuum limit coincides
with that of either Euclidean or causal dynamical triangulations in two dimensions. 
It would be exciting if LCDT constituted a {\it new} universality class, but this seems a priori less 
likely because of the apparent scarcity of universality classes among two-dimensional models of pure
geometry without matter coupling, 

If LCDT quantum gravity were to lie in one of the two known universality of DT models in two dimensions,
our best guess at this stage would be that it is equivalent to CDT, for two
reasons: first, there is good evidence that this is  true in three spacetime dimensions \cite{jordanloll},
in the sense that there one finds in both models a phase whose ground state has the scaling
properties of a Euclidean de Sitter universe. (Of course, this by no means constitutes a proof that the same happens
in two dimensions, which differs in terms of both its geometric degrees of freedom and its phase structure.) 
Second, the difference between DT and CDT in two dimensions has so far been explained
in terms of the absence of baby universes in the latter \cite{alet,ackl}.\footnote{More precisely, it is the absence
of the possibility for baby universes to proliferate without limit; a limited, controlled presence of baby universes
{\it is} compatible with two-dimensional CDT, as demonstrated by the model of generalized CDT \cite{sft-cdt,ambjornbudd}.} 
Since the condition of vertex causality in 
locally causal DT suppresses the light cone degeneracies characteristic of topology change and therefore of the
creation of baby universes, the LCDT model seems closer to CDT than to DT, where baby universes dominate.

\subsection{Volume profile}
\label{vol:sec}

We begin by examining the so-called volume profile of a typical geometric configuration generated by
the Monte Carlo simulation of LCDT. The volume profile is simply given by the size of the
spatial volume as a function of time. Since by construction the model does not have a distinguished
notion of time in terms of some lattice substructure, we will make use of the notion of time
introduced in Sec.\ \ref{ctc-sec} above. For this purpose, a spatial slice at fixed time is any spacelike cycle -- as
defined at the beginning of Sec.\ \ref{ctc-sec} -- and its time label is obtained by averaging over the
time labels of all of its vertices. The volume of a spatial slice is the number of links contained in it.
To determine the complete volume profile of a spacetime configuration, one has to identify all of its
spatial cycles, which is a nontrivial task. Recall  that unlike in CDT, in LCDT different spatial cycles can 
cross and overlap along some subset of spatial edges. 
\begin{figure}[htb]
\centering
\scalebox{0.65}{\includegraphics{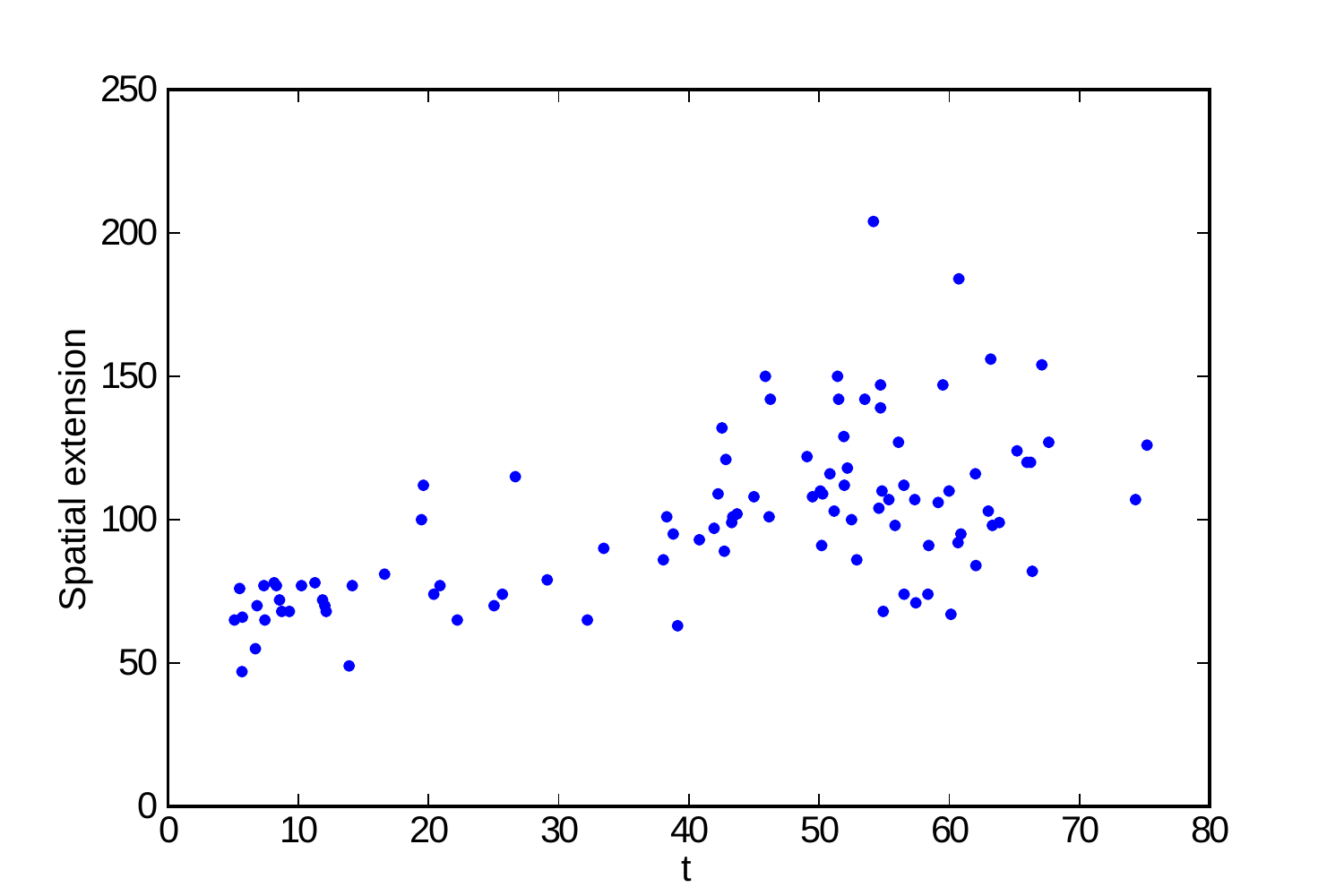}}
\caption{Sampled LCDT volume profile for $N_2\! =\!10.000$, where $t$ denotes the time label of a spacelike cycle and the 
discrete spatial extension is the number of spacelike links in the cycle at a given $t$. 
}
\label{volprof}
\end{figure}

For simplicity, we have considered sampled volume profiles instead of complete ones; for a geometry of 
volume $N_2\! =\! 10.000$ we have randomly sampled 100 spatial slices, and for each slice
determined its volume and time label. Fig.\ \ref{volprof} shows one such sample, to illustrate the situation. We
note that some time labels are very close to each other, which indicates that they probably share one or more
spacelike links. The sample shows large volume fluctuations within small
time intervals, without any discernible overall shape.
This finding is only qualitative, but it is compatible with the typical, strongly oscillating volume profiles
encountered in simulations of two-dimensional CDT \cite{cdtmatter1}.

\subsection{Behaviour of minimal loops}
\label{loop:sec}

A phenomenon that will potentially affect the measurement of dimensions discussed below is the overall shape of
the toroidal configurations. We saw in the previous subsection that their volume profiles seem to be strongly
fluctuating, and are comparable to what one finds in two-dimensional CDT quantum gravity. 
Large fluctuations are commonplace in two dimensions, because there is only a single coupling constant 
(the cosmological constant) which
sets the scale of both the spatial ``universe" and its quantum fluctuations \cite{al}.\footnote{The unique length scale
of the quantum theory is 
$\Lambda^{-1/2}$, where $\Lambda$ is the dimensionful, renormalized cosmological constant.} This is in line with
the fact that general relativity in two dimensions is trivial. 

A new feature in LCDT is the variable length of
the configurations in the time direction, since by construction the fixed time slicing is absent. 
As a result, both the spatial extension of the universe and its time extension -- determined by the prescription 
used for measuring the volume profile, say -- will evolve dynamically. 
For example, the torus may become very thin in one of its directions, an effect which may
be quantified by monitoring the length of closed non-contractible curves. While in CDT individual slices of constant
time can become very short (the minimal length of a spatial $S^1$ compatible with manifold conditions is attained by 
cycles of three links), the probability for this to happen can be made very small by choosing the total
time extent $t_{TOT}$ and the total volume $N_2$ suitably. By contrast, 
in LCDT it can in principle happen that the torus becomes {\it uniformly} thin in one of its directions, even for 
large $N_2$, if this is dynamically preferred.

The relevance of this for the measurement of dynamical dimensions is the possible appearance of finite-size
effects, even when the total volume is large. For example, this happens when the paths of random walkers -- used 
to determine the spectral dimension -- start winding around the torus more than once. To obtain an estimate of
the size of this effect we have set up an algorithm which searches the tori for minimal non-contractible loops. 
It does not distinguish between time- and spacelike links, which implies that the minimal loops found
can be made up of any link types. 

\begin{figure}
\centering
\scalebox{0.55}{\includegraphics{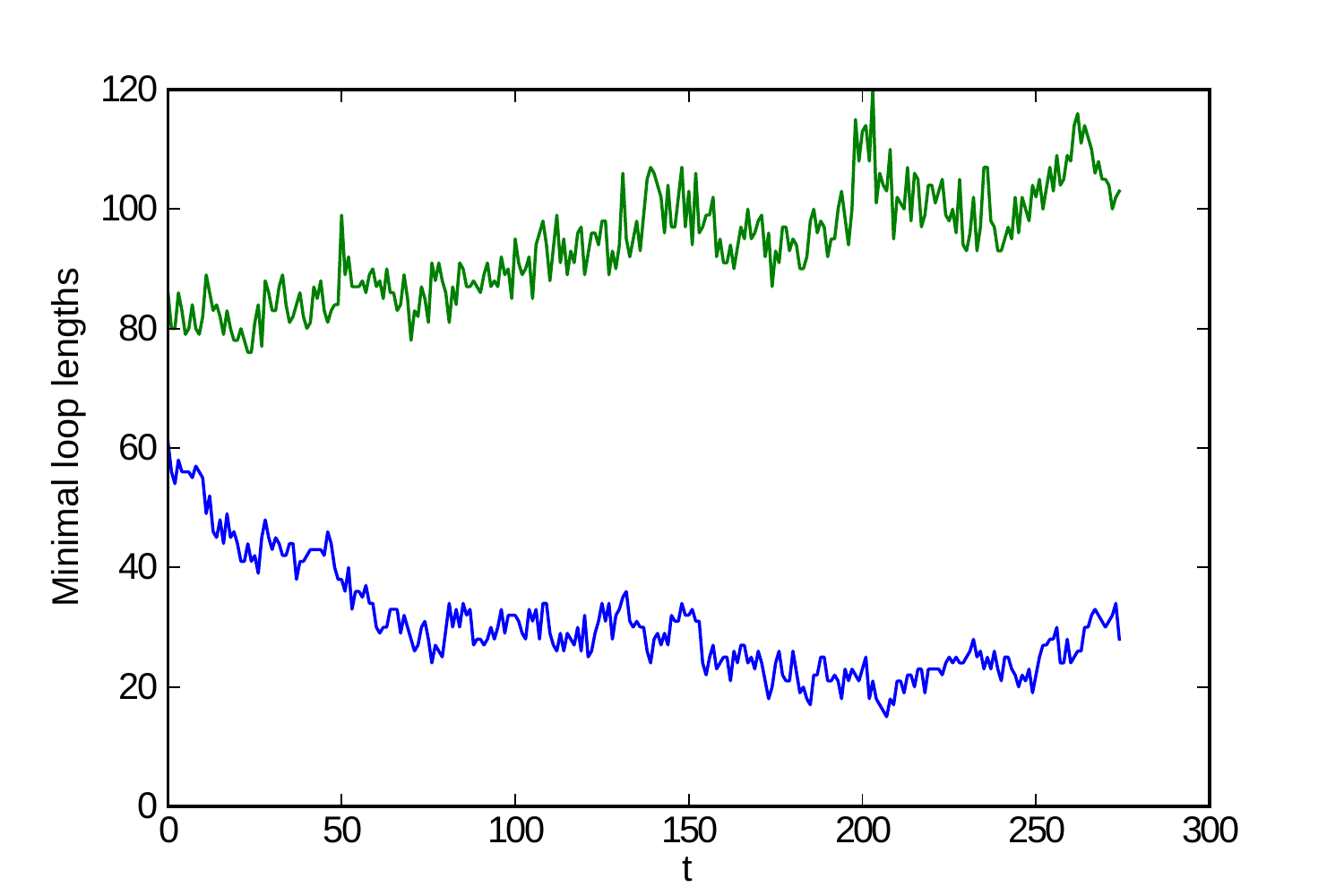}}
\caption{Evolution of the shortest and longest loop lengths
from a sample of minimal non-contractible loops 
through randomly chosen vertices on a given triangulation, as a function of
Monte Carlo time $t$, and at discrete volume $N_2\! =\! 100.000$. 
Largest $t$-values are just before the onset of thermalization. 
}
\label{loop_bounds}
\end{figure}
The algorithm consists of the following steps: 
on a given LCDT configuration $T$, pick a vertex $v$. Starting at $v$, perform a breadth-first search until 
the area searched starts to self-overlap at some other vertex $v'$. Next, determine whether the 
minimal closed curve $c$ through $v$ and $v'$ obtained in this way is contractible or not, by starting
breadth-first searches on either side of $c$. If those searches start overlapping, $c$ is a non-contractible closed 
curve through $v$ of minimal length and we record its length. 
Because this procedure is rather costly in computational terms, we do not repeat it for every vertex of
$T$, but only for a sample of 1.000 randomly chosen initial vertices on $T$. For a given triangulation $T$
we therefore  end up with a sample of locally minimal loops (``locally" because they pass
through prescribed vertices on the torus). 

To obtain Fig.\ \ref{loop_bounds}, we have performed the
sampling of minimal loops at each step during the thermalization of a Monte Carlo simulation of 280 time steps,
corresponding to 280 sweeps, for a triangulation size $N_2\! =\! 100.000$. At each step, we plot only the
shortest and longest minimal loop length of the sample; all other minimal loop lengths lie in between these
two values. 
Since our samples are quite large, the lower curve is probably a good indicator of
the global minimum of the length of non-contractible loops on $T$, which roughly speaking lies in the range 
20--35. This is far away from the kinematically allowed minimum of 3
and shows that the torus does not become
very thin in some places.\footnote{Note that our algorithm does not determine which of the torus directions 
any particular minimal loop winds around. Also in this
respect the information is distinct from that contained in volume profiles of the kind shown in Fig.\ \ref{volprof}.}
In addition, the fact that there is a significant distance between the upper and lower curves shows that the
torus does not degenerate by becoming very long in one direction and uniformly short in the other. 
On the other hand, a triangle count of $N_2\! =\! 100.000$ according to (\ref{pilcdt}) at $\alpha\! =\! 1$
corresponds to a volume $V\! =\! \sqrt{3} N_2/4\! \approx\! 43.300$, where we have set the lattice constant
to unity, $a\! =\! 1$. Assuming the two torus directions are approximately of equal length, this
corresponds to an average linear extension of the torus in the range of 150-200, of which the shortest
loop length therefore is only a small fraction. It implies that fluctuations in the linear extension of the LCDT configurations
are {\it large}, even if the total volume is also large. We will comment further on this characteristic feature
of two-dimensional quantum gravity in Sec.\ \ref{haus:sec} below. 

The overall conclusion is that at system size $N_2\! =\! 100.000$ finite-size effects for observables involving shortest
(geodesic) distances should not 
play a role at least up to link distances of about 30, and for observables involving closed random walks 
at least up to about 500 steps.\footnote{Note that we are not making a distinction between link distance
on the triangulated lattice and link distance on the dual lattice. We have not determined the relative scale between
these two notions of geodesic distance on typical LCDT configurations. The numbers given in the text should 
therefore be treated only as rough estimates.}
Looking ahead to the measurement of dimensions presented below, this still
leaves plenty of room for finite-size effects on larger scales, but they are not quantified easily 
above the thresholds just mentioned.

\subsection{Spectral dimension}
\label{spec:sec}

Dynamical dimensions, like the spectral and Hausdorff dimension, are important and popular examples of observables
in models of nonperturbative quantum gravity because of their computational accessibility in many different
contexts.  
A key insight is that the values of these dimensions do not have to coincide with the dimensionality of the
triangular building blocks used to construct the regularized model if one takes a nontrivial, infinite
continuum limit, as we are doing. Furthermore, a familiar feature from studying graphs and fractals, namely, the fact
that there exist ``spaces" of non-integer dimension, is also encountered in systems of dynamical triangulations.
This is not necessarily inconsistent from a physical point of view as long as the anomalous values of the
dimensions are confined to a highly quantum-fluctuating, non-semiclassical regime, 
typically at the Planck scale. Since there is
no nontrivial classical theory of two-dimensional general relativity whose solutions might be recovered from a
corresponding quantum theory in the limit as $\hbar\rightarrow 0$, 
there are no a priori physicality constraints on the Hausdorff and spectral dimension of
an ensemble\footnote{When talking about ``the Hausdorff dimension", say, of (C)DT, we mean of course 
the {\it expectation value} of this quantity, measured in the ground state of the relevant ensemble.} 
of DT configurations in two dimensions, causal or otherwise. 

For Euclidean DT in two dimensions, from theoretical scaling arguments
the spectral dimension is 2 and the Hausdorff dimension 4 \cite{dim2d}, which has also been
corroborated numerically \cite{dim2dnum}. Invoking an equivalence
between CDT configurations and tree graphs, CDT in 1+1 dimensions can be shown to
have a spectral dimension of at most 2 and a Hausdorff dimension of almost surely 2 \cite{djw},
the latter in agreement with earlier theoretical \cite{al,alet} and numerical \cite{cdtmatter1} results. 
In the context of LCDT, 
we will first investigate the spectral dimension. In the section following this one, we will examine 
the Hausdorff dimension, which appears to
be the quantity best suited to discriminating between the different universality classes.

\begin{figure}
\centering
\scalebox{0.55}{\includegraphics{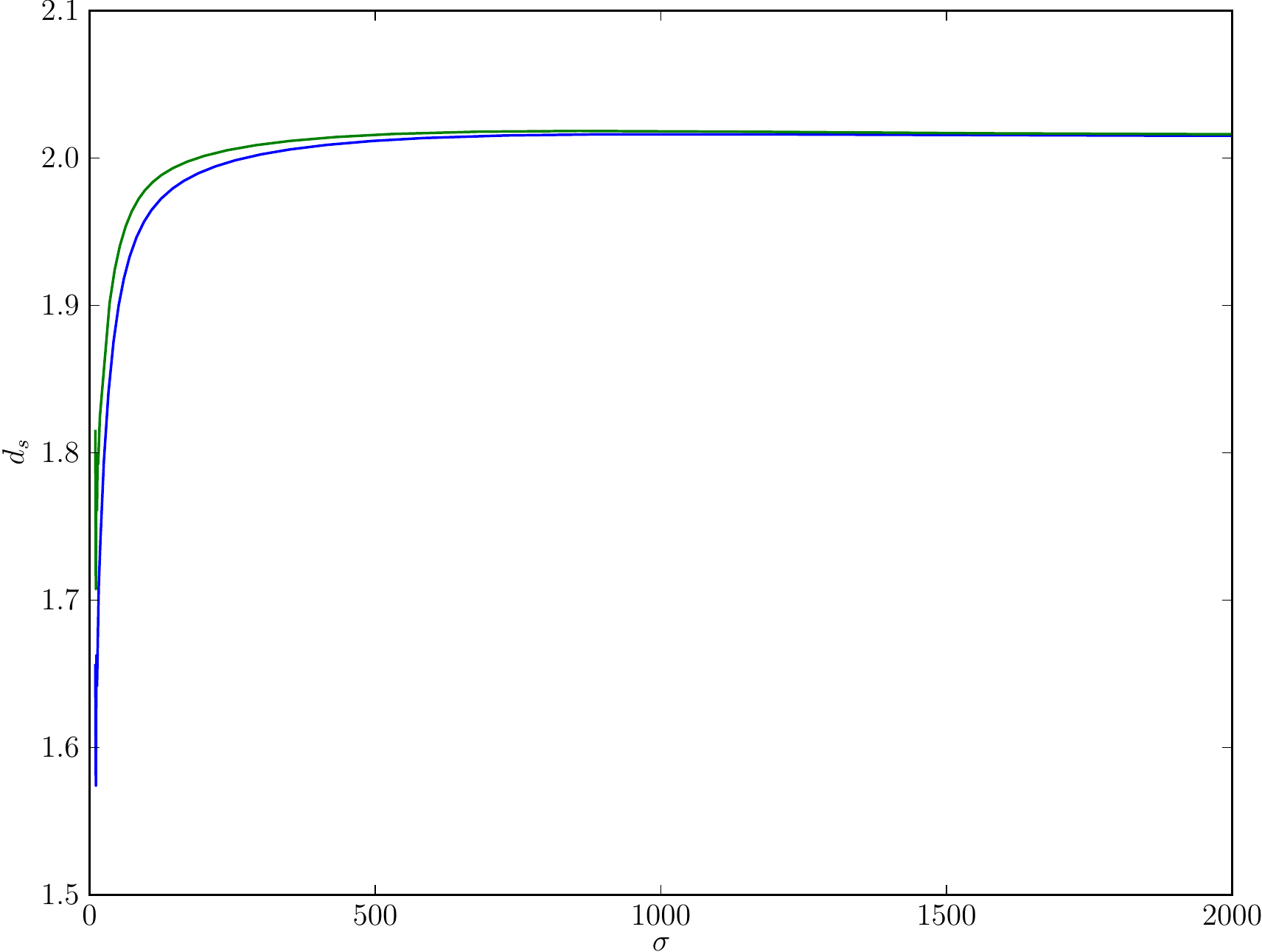}}
\caption{The spectral dimension of CDT as a function of the diffusion time $\sigma$, measured at volume 
$N_2\! =\! 100.000$, and time extension $t_{TOT}\! =\! 80$. 
The upper, green line is the dimension $d_s^{(1)}$ of eq.\ (\ref{canon_disc}) and the lower, blue line is
the dimension $d_s^{(2)}$ of eq.\ (\ref{disc}). 
Statistical error bars are too small to be displayed.}
\label{spec_cdt}
\end{figure}

\begin{figure}
\centering
\scalebox{0.55}{\includegraphics{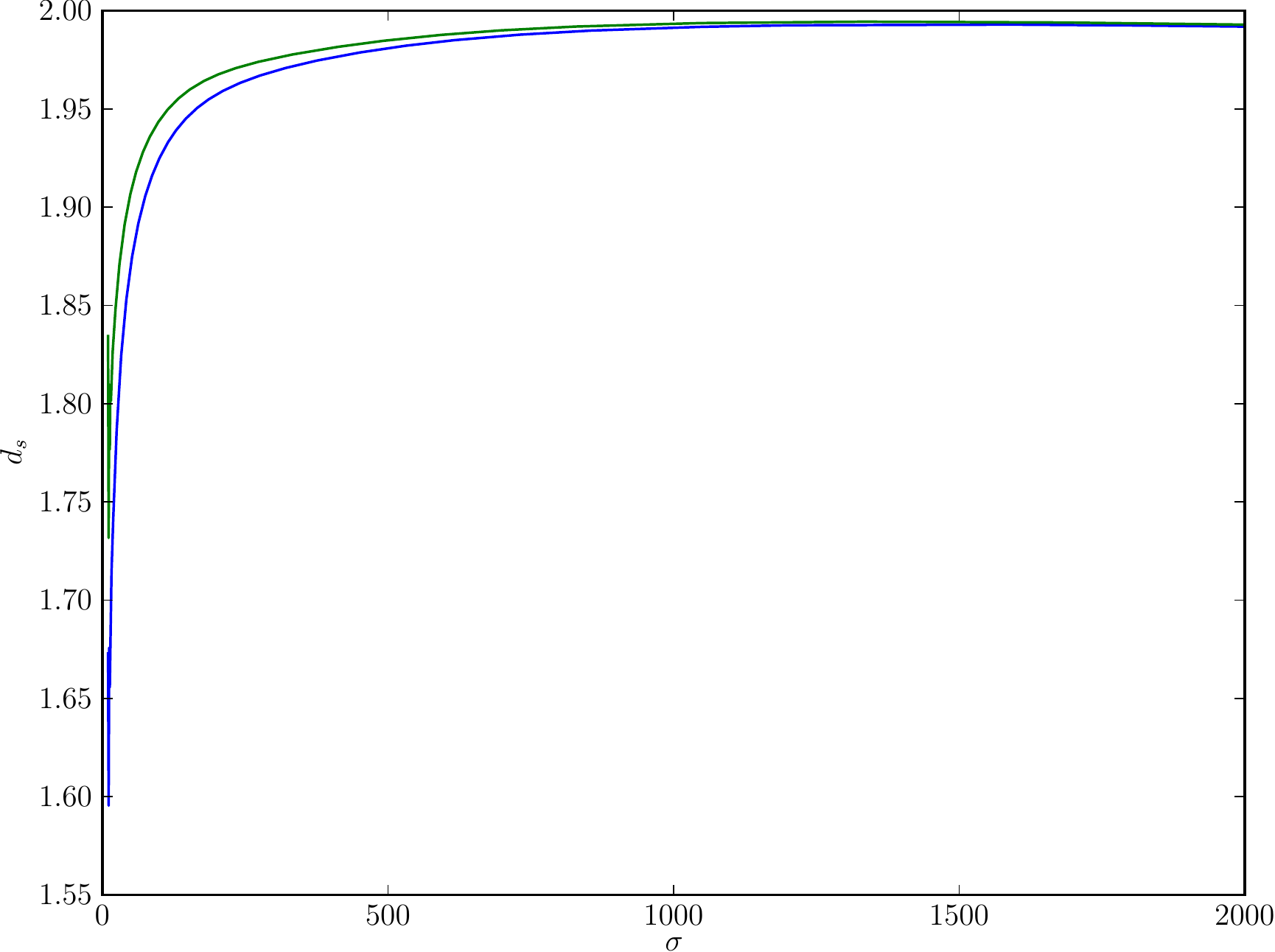}}
\caption{The spectral dimension of locally causal DT (with $\alpha\! =\! 1$) as a function of the diffusion time $\sigma$, 
measured at volume $N_2=100.000$. 
The upper, green line is the dimension $d_s^{(1)}$ of eq.\ (\ref{canon_disc}) and the lower, blue line is
the dimension $d_s^{(2)}$ of eq.\ (\ref{disc}). 
Statistical error bars are too small to be displayed.}
\label{spec_iso}
\end{figure}

The first step in measuring the spectral dimension is to define a discrete diffusion process 
on a two-dimensional Euclideanized locally causal triangulation $T$. This takes the form of a random 
walk moving in steps of unit distance between the centres of neighbouring triangles as function
of a discrete external diffusion time $\sigma$, analogous to what was done
in CDT in four dimensions \cite{reconstruct,spectral}. In other words, the diffusion takes place along the edges of
the trivalent lattice dual to $T$. Calling $K_T (i,i_0;\sigma)$ the probability to go from triangle $i_0$
to triangle $i$ in $\sigma$ steps, satisfying $\sum_i K_T (i,i_0;\sigma)=1$,
the discrete diffusion equation on the triangulation $T$ reads
\beq
\label{diffu}
K_T(i,i_0;\sigma+1) =  (1-\chi) K_T (i,i_0;\sigma) + \, \frac{\chi}{3} \sum_{j\, {\rm n.n. \, of}\, i} K_T (j,i_0; \sigma),
\eeq 
subject to the initial condition $K_T(i, i_0;\sigma\! =\! 0) = \delta_{i,i_0}$.
The sum on the right-hand side of (\ref{diffu}) is over the three nearest neighbours $j$ of triangle $i$,
and $\chi\in [0,1]$ is a diffusion constant which allows for a non-vanishing probability $(1-\chi)$ that the random
walker remains at the same triangle during a diffusion step. 
It is included merely for convenience, to somewhat smoothen out the discretization artefacts for short diffusion paths.
In particular, there is an asymmetry between paths of even and odd numbers of steps (c.f. the discussion 
in \cite{reconstruct}), with a corresponding oscillatory behaviour in the curve for $d_s$ that is also
present in our Figs.\ \ref{spec_cdt} and \ref{spec_iso} when one zooms into the region below
$\sigma \approx 50$. A diffusion constant $\chi\! <\! 1$ 
has been used previously when studying the spectral dimension in three-dimensional CDT \cite{benehenson}.
In our simulations, we have worked with $\chi\! =\! 0.8$ throughout.

To extract the spectral dimension, we consider closed random walks, beginning and ending at a
specified triangle $i_0$. They enter into the calculation of the average return probability 
\beq
P_T(\sigma)= \frac{1}{N_2(T)}\sum_{i_0 \in T} K_T(i_0,i_0;\sigma)
\label{arp}
\eeq
for a given triangulation $T$. The spectral dimension $d_s$ is obtained from the expectation value 
$\langle P(\sigma )\rangle_{N_2}$ of the observable (\ref{arp}) in the ensemble of triangulations of fixed
volume $N_2$ according to
\beq
\label{diffdim}
d_s(\sigma) := -2 \,\frac{\operatorname{d} \ln \langle P(\sigma)\rangle}{\operatorname{d} \ln \sigma }.
\eeq
In practice, we perform a ``double" random sampling, where for each randomly chosen triangulation
we pick 10 times a triangle randomly as starting point $i_0$ for a random walk, and
then repeat the process for at least 400 triangulations.
For a diffusion process on classical, flat $\R^d$, the formula analogous to (\ref{diffdim}) 
simply reproduces the topological dimension $d$, independent 
of $\sigma$, but in the quantum context the behaviour of $d_s$ can be 
more complicated and, generally speaking, $\sigma$-dependent. 
By not denoting the $N_2$-dependence in (\ref{diffdim}) explicitly we mean to indicate that $d_s$
is determined in the limit of large volumes where this dependence gradually disappears.

When the total spacetime is compact, the spectral
dimension will always go to zero for sufficiently large $\sigma$. This is a finite-size effect which occurs when 
generic random walks become sufficiently long to wrap around space one or more times. 
We are primarily interested in the $\sigma$-regime below this range.
If the system develops a stable plateau below the scale where significant
finite-size effects kick in, we will refer to this constant value of
$d_s$ as the spectral dimension of the underlying ``quantum spacetime". Note that if the system size is too small,
a plateau will never form due to a dominance of finite-size effects.

Since there are no published numerical results on the spectral dimension of two-dimensional CDT, and since it will be 
useful to have a point of reference for the measurements in LCDT, we will first present our results for
the spectral dimension of regular CDT quantum gravity on a two-torus. As explained earlier, the reduction to
pure CDT configurations is achieved by setting $\alpha\! =\! 1/4$ in the action.
To understand better the effects
of the discretization, we have used two different discrete versions of the defining formula (\ref{diffdim}) for
the spectral dimension. Having determined the expectation value $\langle P(\sigma)\rangle$ for integer $\sigma$
from the data, we have employed two different implementations in terms of finite differences. 
The standard choice is
\beq
d_s^{(1)} = -2\, \frac{\ln \langle P(\sigma + 1)\rangle - \ln \langle P(\sigma)\rangle }{\ln(\sigma + 1) - \ln \sigma }.
\label{canon_disc}
\eeq
In addition, we have used the alternative form
\beq
d_s^{(2)}  = -2 \sigma \left(\frac{\langle P(\sigma + 1)\rangle }{\langle P(\sigma)\rangle } - 1 \right),
\label{disc}
\eeq
which has the same continuum limit and the advantage that no expensive functions are required.

The CDT results for the spectral dimension $d_s$ are displayed in Fig.\ \ref{spec_cdt}, for data taken at volume
$N_2\! =\! 100.000$ and time extension $t_{TOT}\! =\! 80$. 
For $\sigma \lesssim 700$ there is a small discrepancy between the curves corresponding to
the two different discretizations, giving us an estimate of the systematic error of determining $d_s$ for small
values of $\sigma$. For larger $\sigma$, both curves merge into what is essentially a single plateau. 
The spectral dimension of CDT extracted from data on the plateau is $d_s = 2.02 \pm 0.02$, 
in good agreement with the expected value of 2. 

The curves for the spectral dimension for LCDT are shown in Fig.\ \ref{spec_iso}. 
Qualitatively the plot is similar to that of
CDT, but the plateau is reached only for somewhat larger diffusion times $\sigma \gtrsim 1.000$. 
The numerical result for the spectral dimension is $d_s=1.99 \pm 0.02$, which we regard as a
convincing confirmation that the spectral dimension of locally causal DT is 2, like that for DT and CDT
quantum gravity.

\subsection{Hausdorff dimension}
\label{haus:sec}

The Hausdorff dimension $d_h$ is a key quantity to discriminate between distinct universality classes
of two-dimensional DT quantum gravity. The general idea is to relate the volume $V$ of compact, connected regions 
in space -- typically discrete analogues of geodesic balls around a chosen point --
to their linear size, e.g. the radius $r$ of the region, and to extract the leading scaling behaviour from 
$\langle V(r)\rangle\! \sim\! r^{d_h}$. For our purposes, we will use a ``differential" version of this relation, where one monitors
the one-dimensional volumes of spherical shells around a given triangle $i_0$ or, equivalently, the number of (dual) vertices
at radial distance $r$ from a vertex $i_0$
of the lattice dual to a given LCDT configuration $T$. We define $n(r,i_0)$ as the number of triangles found at geodesic
distance $r$ from $i_0$, where geodesic distance is defined as the (integer) length of the shortest path along edges of
the dual lattice. We have $n(0,i_0)\! =\! 1$ and $n(1,i_0)\! =\! 3$ for all $i_0$, because each triangle has exactly three
neighbours and therefore the dual lattice is trivalent. Every triangle of $T$ will appear in exactly one of the shells, 
implying that $\sum_r n(r,i_0)\! =\! N_2(T)$.
The identification of the shells can be implemented as a modified breadth-first search, which keeps track of
when a change of shells occurs.
Averaging over all initial triangles $i_0\in T$, we obtain the average shell volumes at radius $r$,
\begin{equation}
\label{avshell}
n(r) = \frac{1}{N_2} \sum_{i_0=1}^{N_2} n(r, i_0).
\end{equation}
In what follows, we will refer to the function $n(r)$ as the {\it shape} of a triangulation. 

To extract the
Hausdorff dimension, we have applied finite-size scaling to the expectation value $\langle n(r)\rangle$ of
the shape function (\ref{avshell}).\footnote{We also tried to extract the Hausdorff dimension from the
scaling relation ${\bar r}(N_2)\sim N_2^{1/d_h}$ for the average linear extension $\bar r$ defined in
eq.\ (\ref{avlin}), but this did not yield meaningful results because of the convergence issues to be described
in more detail below.} 
The simulations consisted of 48.000 sweeps each at volumes $N_2=100.000$, $200.000$,
$300.000$ and $400.000$, and were done for LCDT (at $\alpha\! =\! 1$) and, for reference and
comparison, also for CDT (corresponding
to $\alpha\! =\! 1/4$). The scaling ansatz for the radius and shell volume is 
$r \rightarrow x=N_2^{-1/d_h}r$ and $\langle n(r)\rangle \rightarrow N_2^{-1 + 1/d_h} \langle n(r)\rangle $ respectively.
Individual data points $\langle n(r)\rangle$ were transformed into curves via spline interpolation and a
Levenberg-Marquardt least-square fit was used to align the shapes \cite{ruijl}.

\begin{figure}[htb]
\centering
\scalebox{0.65}{\includegraphics{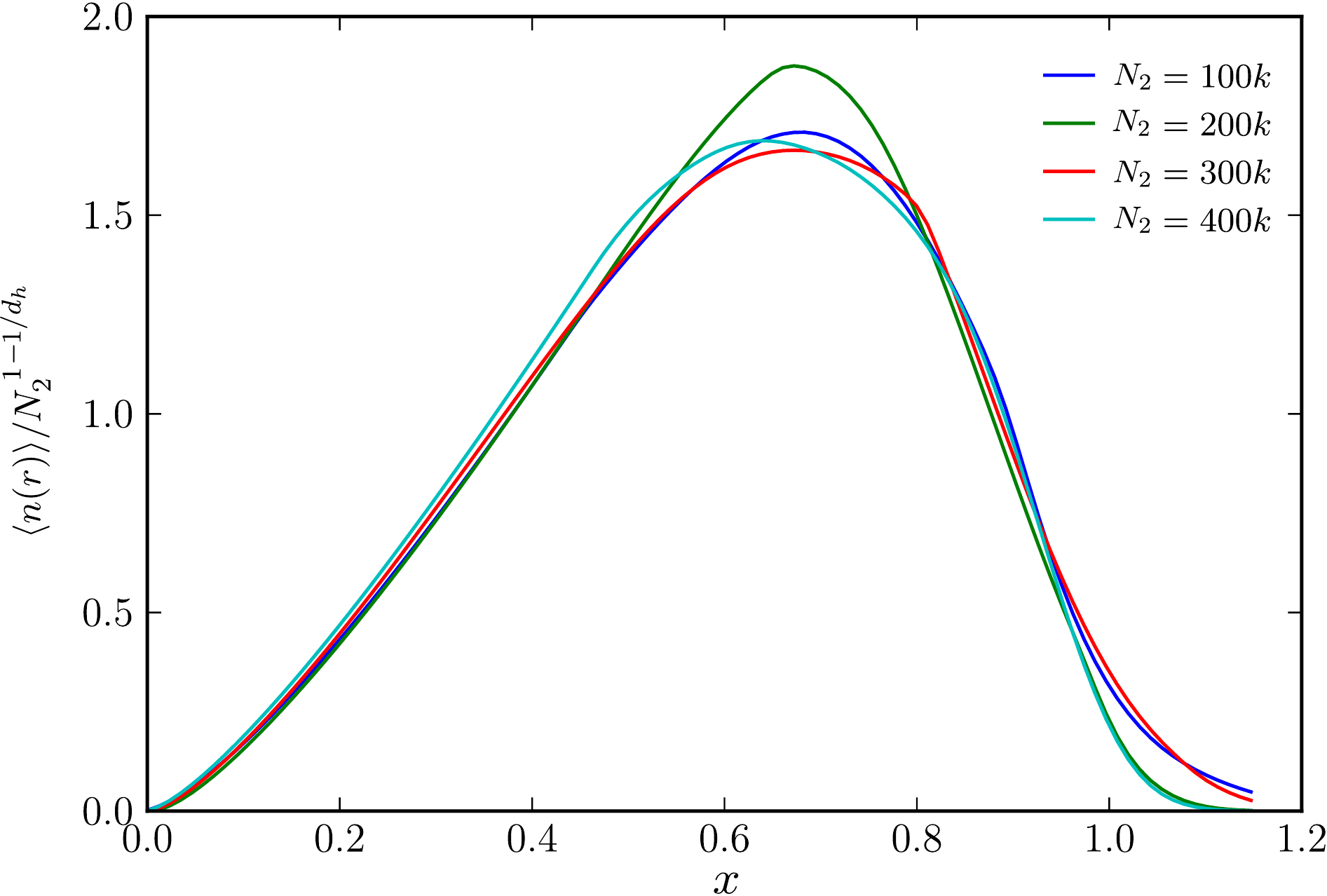}}
\caption{CDT quantum gravity: Fit for best overlap of the rescaled shapes $\langle n(r)\rangle /N_2^{-1+1/d_h}$ 
as function of the rescaled 
distance $x\! =\! r/N_2^{1/d_h}$, for Hausdorff dimension $d_h\! =\! 2.2$. 
The extension in time direction was set to $t_{TOT}\! =\! 80$.}

\label{haus_curve_fit_cdt}
\end{figure}

In maximizing the overlap we have taken into account all $x$-values where $\langle n(r)\rangle$ 
has at least half of its maximal value. 
We have also measured the ``short-distance'' Hausdorff dimension for small $x$ by optimizing the overlap of 
the initial rising slopes of the curves $\langle n(r)\rangle $. In principle this dimension need not coincide with the global 
Hausdorff dimension $d_h$ we have been considering \cite{cdtmatter2}, but in our case there was little difference.
Our results for the best overlap of the shape functions for the CDT case are shown in Fig.\ \ref{haus_curve_fit_cdt}; 
they correspond to a Hausdorff dimension $d_h=2.2 \pm 0.2$, which is compatible with the known 
value of 2.\footnote{For CDT in two dimensions, the Hausdorff dimension has been measured 
previously \cite{cdtmatter1},
with good results, from finite-size scaling of the distribution of spatial volumes. Because of the absence
of a pre-defined time function, this method is not computationally feasible for LCDT.} Our
error bars are rather large, because the best fit depends quite sensitively on the $x$-range for which the
overlap is optimized.

\begin{figure}[htb]
\centering
\scalebox{0.65}{\includegraphics{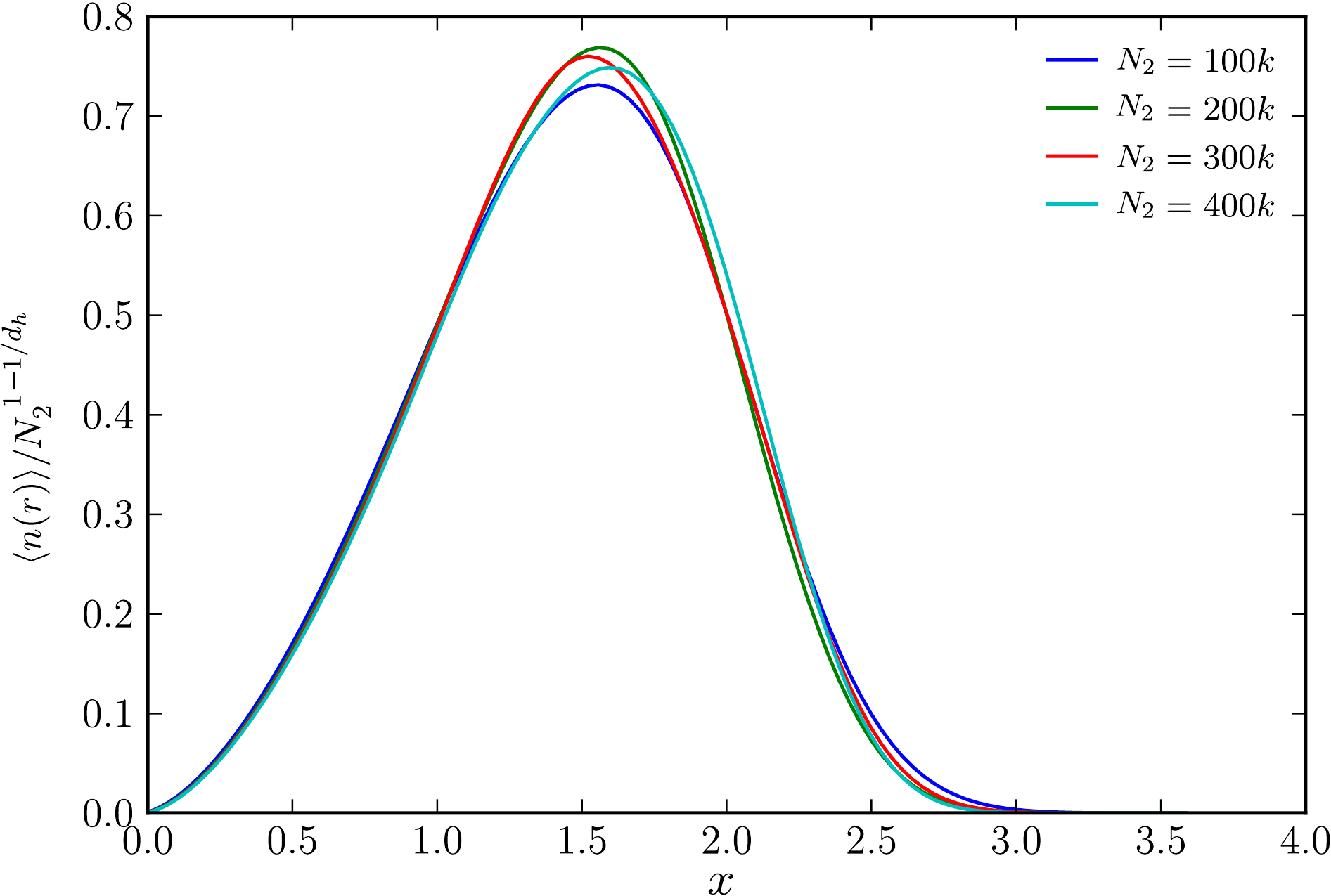}}
\caption{LCDT quantum gravity: Fit for best overlap of the rescaled shapes $\langle n(r)\rangle /N_2^{-1+1/d_h}$ 
as function of the rescaled 
distance $x\! =\! r/N_2^{1/d_h}$, for Hausdorff dimension $d_h\! =\! 2.7$.}
\label{haus_curve_fit}
\end{figure}

The analogous data for LCDT are displayed in Fig.\ \ref{haus_curve_fit}. Maximal overlap is achieved for
a Hausdorff dimension $d_h\! =\! 2.71 \pm 0.2$, which is far away from our conjectured CDT value of 2,
and even further away from the DT value of 4. We conclude that LCDT very likely lies not in the same
universality class as DT. Contrary to our expectation, equivalence of LCDT and CDT appears to be excluded too. 
Instead, our measurements point towards LCDT lying in
a {\it new} universality class, not hitherto seen in quantum models of two-dimensional pure gravity. 
This would be a truly interesting result, and it warrants another critical look at the strength of our evidence.    

As is apparent from Figs.\ \ref{haus_curve_fit_cdt} and \ref{haus_curve_fit}, the quality of the overlaps 
is not very good. Could there be systematic sources
of error that affect our results to the extent that they ultimately are {\it not} in contradiction with $d_h\! =\! 2$ for
LCDT? In other words, may we be underestimating our error bars significantly? 
It may be worth recalling that it took some time to nail down the 
Hausdorff dimension of two-dimensional DT quantum gravity numerically. 
In the words of the authors of \cite{catt_thor}, early simulation results were 
``remarkably inconclusive" (see \cite{catt_thor} for 
further references). The same work also used finite-size scaling
of the shape function to determine $d_h$, with a fit quality somewhat similar to ours. Especially when using the
dual lattice -- as we are also doing in the present work -- the Hausdorff dimension extracted this way was significantly
off the mark ($d_h\! =\! 3.15$ instead of the known, correct value 4). Of course, one should keep in mind
that these simulations were 
performed for a geometric ensemble different from LCDT and for moderate
lattice sizes $N_2\leq 32.000$ only. On the other hand, the causal gluing rules of LCDT introduce a local
``stiffness" in the configurations compared to DT, which is likely to require larger volumes to achieve 
numerical results of comparable quality. 

In the case of DT simulations, significant progress with respect to 
the convergence of fits was obtained by introducing 
a ``phenomenologically fudged" scaling relation for the geodesic distance \cite{dim2dnum}, namely,
\begin{equation}
\label{fudge}
x=\frac{r+a}{N_2^{1/d_h} + b},
\end{equation}
where $a$ and $b$ are two parameters meant to compensate lattice artifacts at short distances. 
We have also tested relation (\ref{fudge}), 
but found that nonvanishing values for $a$ and $b$ steer the CDT results even further away from 2 and 
also increase the sensitivity to the choice of fitting region.

A generic feature of two-dimensional quantum gravity illustrated by the numerical difficulties 
already mentioned is the fact that in two dimensions quantum fluctuations are always large, even for large
lattice volumes. This is different from CDT in higher dimensions, say, where the dynamics is
governed by two scales:
one macroscopic, related to the overall size of the universe, and another one microscopic, setting the
scale of quantum fluctuations. In two dimensions, there are no nontrivial classical solutions, and 
there is only a single scale, that of the quantum fluctuations. 

In this situation, it is therefore natural for finite-size effects to generically be large,
especially when there are non-contractible
directions along which space can become ``small", as can happen for the torus topology used for 
LCDT.\footnote{Note that the DT simulations mentioned above use
the topology of a two-sphere. Also this indicates the need to go to larger volumes in the LCDT case.}
This is certainly relevant when measuring the Hausdorff dimension; when the geodesic balls
centred at some triangle $i_0$ start wrapping around one of the torus directions, the interpretation of
the scaling relation by which we extract $d_h$ will be affected, in the sense that only triangles not visited
previously will be counted as belonging to a given radial shell. Of course, one can keep extracting
the Hausdorff dimension regardless, but should be aware that it contains also global, topological 
information.

\begin{figure}
\centering
\scalebox{0.7}{\includegraphics{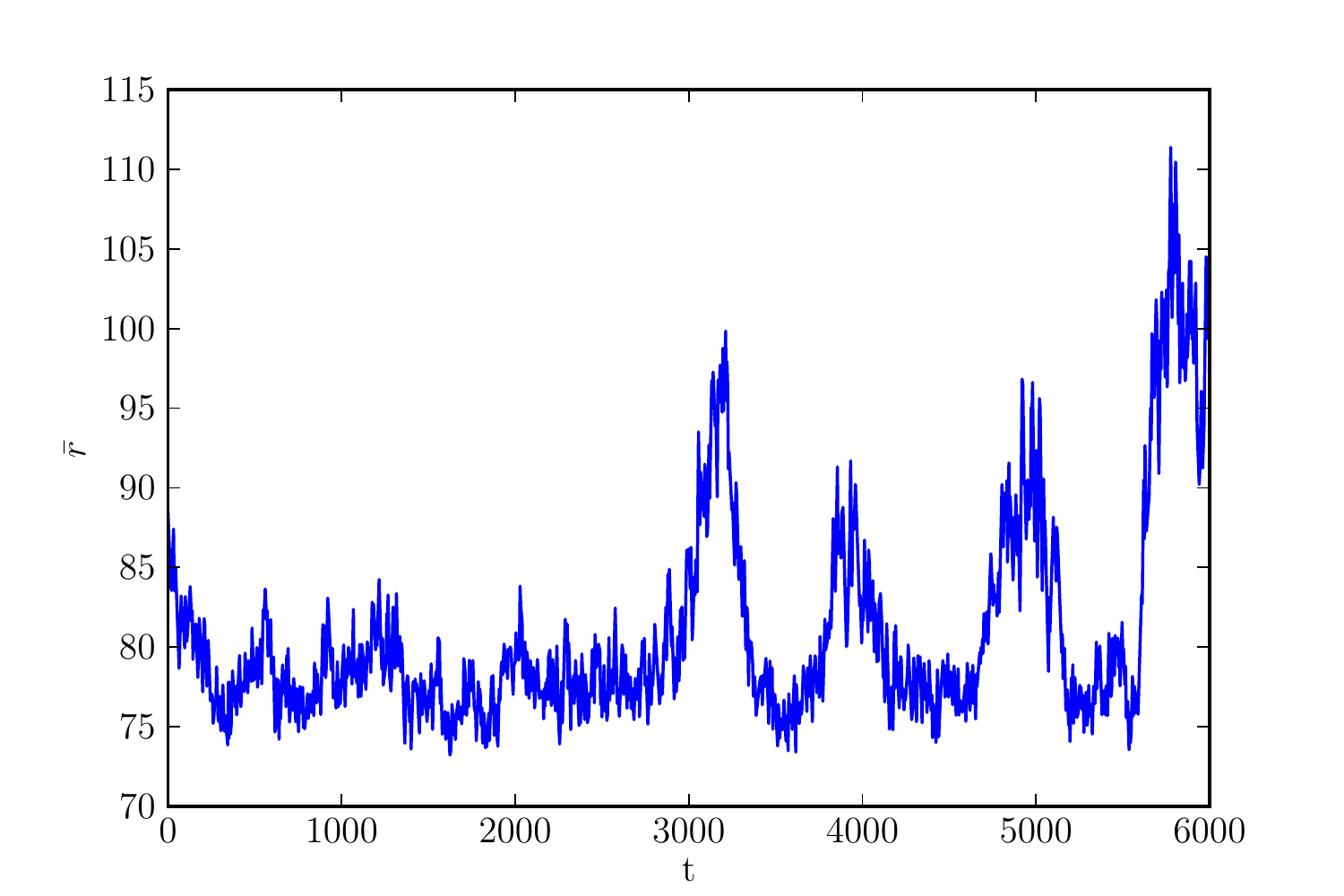}}
\caption{Development in Monte Carlo time $t$ of the average linear extension $\bar{r}$
of a LCDT configuration $T$ of volume $N_2\! =\! 50.000$. Isolated peaks in $\bar{r}$ keep
occurring, even as the number of sweeps becomes very large. (Note that the y-axis has an offset of 70.)}
\label{finsize_haus}
\end{figure}
To get further insights into the origin of the relatively poor quality of our fits, 
we have measured yet another observable, the average linear extension \cite{reconstruct}
\begin{equation}
\label{avlin}
\bar{r} = \frac{1}{N_2} \sum_r r \cdot n(r)
\end{equation}
of a given triangulation $T$ of discrete volume $N_2$, which is just the weighted average of 
the geodesic distance $r$. As is described in more detail in the Appendix, the observable $\bar{r}$
has convergence issues, which appear to persist even on large lattices and after a large number
of sweeps. What seems to happen to the geometrical configurations is that most
of the time they are approximately ``square-shaped", 
with comparable linear extensions for either torus direction, but ever so
often make an excursion to an overall shape that is elongated, where one torus direction becomes
longer than the other one, with $\bar{r}$ increasing as a result. 
After a relative maximum of the two lengths has been reached, the system gradually
reverts back to being square-shaped and stays there for a while before another excursion takes
place (see Fig.\ \ref{finsize_haus} for illustration). 

Of course, square and elongated configurations (for identical volume $N_2$) not only have different
average extensions $\bar{r}$, but also different shape functions $n(r)$ (see Appendix) 
and therefore in general different Hausdorff dimensions. A likely explanation for our inaccurate 
determination of $d_h$ is therefore the failure of the shape to stabilize during the course of the
simulation, and the finite-size effects associated specifically with elongated shapes, in addition to the
already mentioned large magnitude of the quantum fluctuations overall. This is supported by a numerical experiment we
have performed in pure CDT quantum gravity at a volume $N_2\!=\! 9.000$. For $t_{TOT}\! =\! 80$
time slices, the behaviour was ``square-like", in the sense that the
Monte Carlo history of the average extension $\bar{r}$ did not have any peaks. However, when
we shortened the time extension to $t_{TOT}\! = \! 20$, peaks similar to those depicted in Fig.\ \ref{finsize_haus}
appeared.

\begin{figure}[htb]
\vspace{0.5cm}
\centering
\begin{tikzpicture}[scale=0.75]
\pgfmathsetmacro{\s}{sqrt(3)}

\draw[blue] (0,0) -- (\s,1);
\draw[blue] (0,0) -- (\s,-1);
\draw[red] (\s,1) -- (\s,-1);
\draw[blue] (\s,1) -- (4*\s ,1); % upper
\draw[blue] (\s,-1) -- (4*\s,-1); % lower

\draw[red] (\s,1) -- (2*\s,-1);
\draw[red] (2*\s,-1) -- (2*\s,1); % up
\draw[red] (2*\s,-1) -- (3*\s,1);
\draw[red] (3*\s,-1) -- (3*\s,1); % up
\draw[red] (3*\s,1) -- (4*\s,-1);

\draw[red] (4*\s,1) -- (4*\s,-1);
\draw[blue] (4*\s,1) -- (5*\s,0);
\draw[blue] (4*\s,-1) -- (5*\s,0);

\end{tikzpicture}
\vspace{0.3cm}
\caption{A bubble has a $sst$-triangle at either end and arbitrarily many $stt$-triangles in between.}
\label{bubble}
\end{figure}
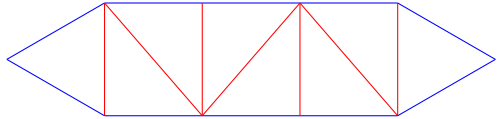

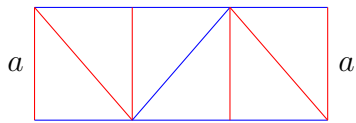
\begin{figure}[htb]
\vspace{0.5cm}
\centering
\begin{tikzpicture}[scale=0.75]
\pgfmathsetmacro{\s}{sqrt(3)}
\draw[red] (\s,1) -- node[black,left] {$a$} (\s,-1);
\draw[blue] (\s,1) -- (4*\s ,1); % upper
\draw[blue] (\s,-1) -- (4*\s,-1); % lower

\draw[red] (\s,1) -- (2*\s,-1);
\draw[red] (2*\s,-1) -- (2*\s,1); % up
\draw[blue] (2*\s,-1) -- (3*\s,1);
\draw[red] (3*\s,-1) -- (3*\s,1); % up
\draw[red] (3*\s,1) -- (4*\s,-1);
\draw[red] (4*\s,1) -- node[black,right] {$a$} (4*\s,-1);

\end{tikzpicture}
\vspace{0.3cm}
\caption{A self-overlapping bubble; the links with label $a$ are to be identified.}
\label{sob}
\end{figure}

Lastly, in our search for ways to improve the convergence behaviour of LCDT quantum gravity, we 
investigated what happens when self-overlapping bubbles are not allowed to occur. We mentioned these
structures briefly in Sec.\ \ref{invest:sec} above. A bubble is a contractible loop of spacelike links, which
in its interior is decorated by timelike links only (Fig.\ \ref{bubble}). It always has two $sst$-triangles at its end points and
consists of $stt$-triangles otherwise. When a bubble winds around a compact torus direction, it can touch
itself again (``self-overlap") along one or more spacelike edges of its boundary (see Fig.\ \ref{sob} for a simple 
example). Self-overlapping bubbles are geometrically significant, because -- depending on their interior
geometry -- they can give rise to timelike cycles as defined at the beginning of Sec.\ \ref{ctc-sec}. Their
appearance is not forbidden by local vertex causality.
Relevant to our present discussion is the fact that 
globally self-overlapping bubbles cause severe thermalization issues in $2+1$ dimensions, and
therefore were removed from the ensemble \cite{jordanthesis,jordanloll}. This motivated us to
remove self-overlapping bubbles from the LCDT ensemble in $1+1$ dimensions too, and to check
whether it makes a difference to the measurement of the Hausdorff dimension.

\begin{figure}[htb]
\centering
\scalebox{0.6}{\includegraphics{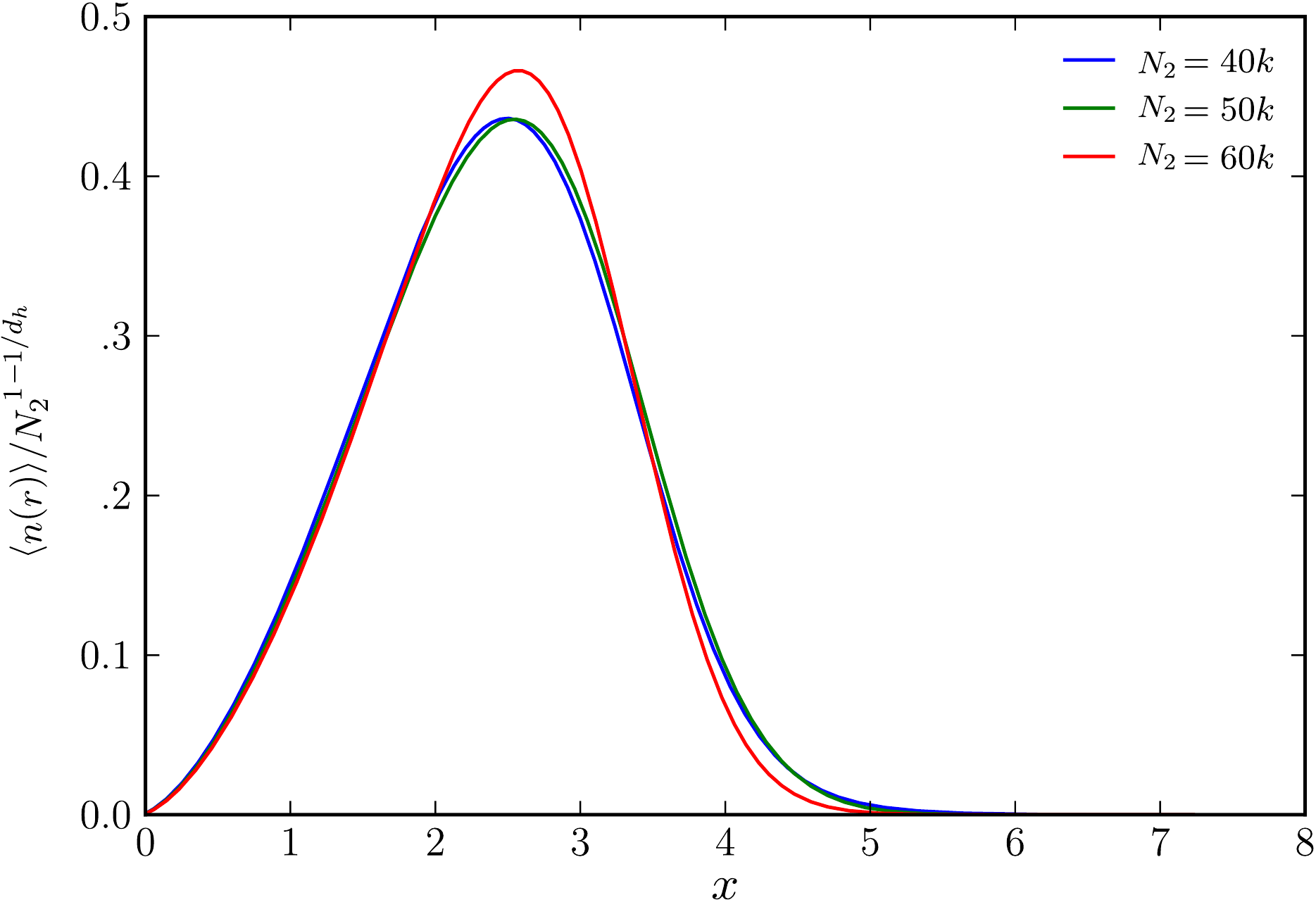}}
\caption{LCDT quantum gravity without self-overlapping bubbles: Fit for best overlap of the rescaled shapes 
$\langle n(r)\rangle /N_2^{-1+1/d_h}$ 
as function of the rescaled 
distance $x\! =\! r/N_2^{1/d_h}$, for Hausdorff dimension $d_h\! =\! 3.1$.}
\label{haus_curve_fit_iso_bubblefilter}
\end{figure}
Detecting whether a self-overlapping bubble is created during the Monte Carlo simulation (and
discarding the corresponding move) is nontrivial, since the property is nonlocal and requires a computationally
expensive walk around the lattice (see \cite{ruijl} for details on implementation). For this reason we
performed the numerical analysis on slightly smaller lattices of volume $N_2\!\leq\! 60.000$. 
Measurement of the average linear extension $\bar{r}$ in this setting,
at $N_2\! =\! 50.000$, still revealed a peak structure similar to that of standard LCDT {\it with} self-overlapping bubbles,
providing evidence that this structure is not responsible for the observed instability.  
Proceeding like before to determine the Hausdorff dimension for this system, via finite-size scaling 
to maximize the overlap in shape (see Fig.\ \ref{haus_curve_fit_iso_bubblefilter}) 
yielded a Hausdorff dimension of $d_h=3.10 \pm 0.2$.  

To summarize, we have pinpointed 
an instability of the system with regard to its global behaviour, due to occasional excursions to a globally
elongated state, which can be observed by monitoring the geometry's average linear extension $\bar{r}$.
This is the likely source of the suboptimal data quality for the measurement of the Hausdorff dimension $d_h$. 
By considering a more general fitting function for extracting $d_h$ and by using a modified ensemble without 
self-overlapping bubbles we have found no hints of additional sources of error or a shift 
of the Hausdorff dimension toward the CDT value of 2.

\section{Conclusions and outlook}
\label{concl:sec}

The aim of our work was to measure observables in locally causal dynamical triangulations in two dimensions,
most importantly, the spectral and the Hausdorff dimensions, and thereby understand the relation
of LCDT to other models of two-dimensional quantum gravity based on dynamical triangulations. 
Our initial hypothesis was that LCDT lies in the same universality class as CDT, where both
spectral and Hausdorff dimension are equal to 2. While our measurement of LCDT's spectral dimension
did yield a value compatible with 2, with only small error margin, this was not true for the
Hausdorff dimension. Although the error bars were significantly larger -- due to an instability 
in the system that persisted even at the largest volumes -- our measurements found 
a Hausdorff dimension of $d_h\! =\! 2.7\pm 0.2$ and $d_h\! =\! 3.1\pm 0.2$ for two slightly different variants of LCDT. 
On the basis of our simulations, it appears that LCDT is not equivalent to either DT or CDT in the continuum
limit. 

This would be an interesting result, because it implies the existence of a new universality class of
two-dimensional quantum gravity in between Euclidean DT and Lorentzian CDT in two dimensions.
The ``in between" could be true quite literally, since within our measuring accuracy the Hausdorff
dimension of LCDT is compatible with 3. A Hausdorff dimension
$d_h\! =\! 3$ has been observed previously, in simulations of CDT quantum gravity in 1+1 dimensions
coupled to eight copies of Ising spins \cite{cdtmatter2} and coupled to several massless scalar fields \cite{cdtscalar}, 
adding some plausibility to the possibility that
a universality class with this property may actually exist. 
Further confirmation of the appearance of this new phase would come from locating 
a phase transition between CDT (corresponding to $\alpha\! =\! 1/4$, at least for fixed volume)
and LCDT as a function of the parameter $\alpha$ in our model. Having already invested
considerable computing resources into the isotropic case $\alpha\! =\! 1$ in the present work,
we leave this investigation to a future publication. --
Needless to say, it would be extremely interesting to
find an analytic solution of the LCDT model, to put our findings on a more definite footing. 

%\vspace{0.3cm}

\subsection*{Acknowledgments} 
We thank J.\ Ambj\o rn for discussion. -- The contribution of RL is part of the 
research programme of the Foundation for Fundamental Research 
on Matter (FOM), financially supported by the Netherlands 
Organisation for Scientific Research (NWO).
\vspace{0.3cm}

\section*{Appendix} 

\begin{figure}[htb]
\centering
\scalebox{0.7}{\includegraphics{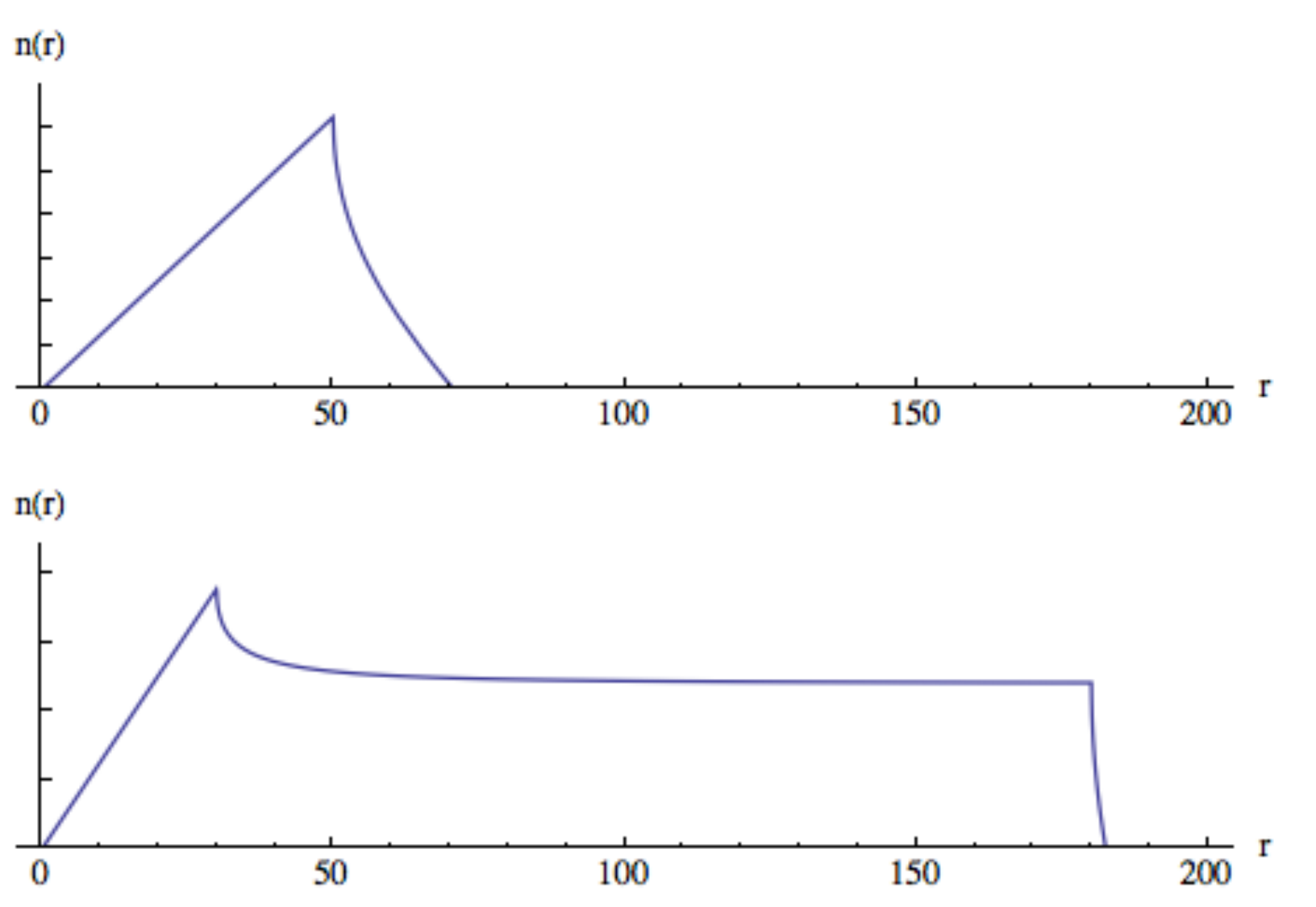}}
\caption{Shape of a flat, classical torus in the continuum: square-shaped of length 50 in either direction (top) and elongated
with extension 30 and 180 in the two directions (bottom).}
\label{toruspeaks}
\end{figure}

In this appendix, we give some more details about the instability we have observed in the LCDT system,
which affects at least one observable, the average linear extension $\bar{r}$ of the universe defined
in (\ref{avlin}), and contributes to the rather poor overlaps we have found in our finite-size scaling analysis
to determine the Hausdorff dimension. Unlike other observables, which typically converge after about
300 sweeps, $\bar{r}$ does not, even for very large system size $N_2\! =\! 400.000$ and after several
thousands of sweeps. To understand better what happens geometrically, we have plotted the shape
$n(r)$ of a typical configuration along the meta-stable ``bottom" of the Monte Carlo history of $\bar{r}$ 
shown in Fig.\ \ref{finsize_haus}, and of a configuration at one of the peaks. 
\begin{figure}[h]
\centering
\scalebox{0.7}{\includegraphics{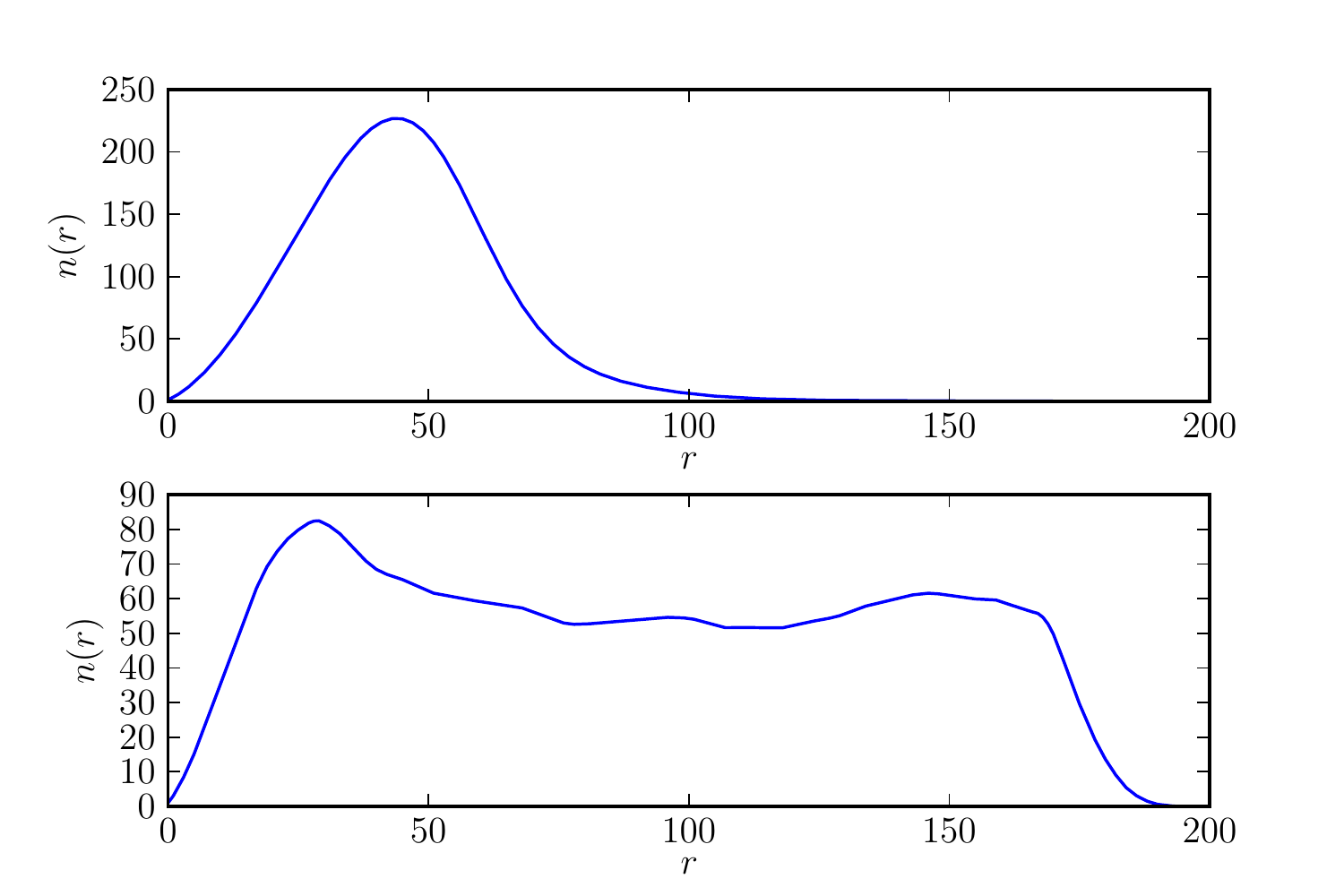}}
\caption{Average shape $n(r)$ (top) and shape at peaks in $\bar{r}$ (bottom) at $N=9.000$, exhibiting a long plateau.}
\label{finsize_haus_3}
\end{figure}
As illustrated by
Fig.\ \ref{finsize_haus_3}, the two are very different. Outside the peaks in $\bar{r}$, the shape of a configuration starts out
with an almost linear increase until it reaches a single maximum, and then quickly drops to zero.
A configuration from a peak in $\bar{r}$ also increases linearly until a first maximum, but then
enters a long plateau before also going to zero. 
These two different shape functions are characteristic for a torus which is approximately
``square-shaped" (i.e. of a similar extension
in either of the torus directions) and one which is elongated. This fact is illustrated by comparing the
measured shapes with those of regular, flat tori\footnote{For simplicity, we are considering only
tori which are obtained from gluing flat rectangles without any ``twists".} in the continuum (Fig.\ \ref{toruspeaks}). 
Despite the totally 
different set-up (single, classical torus without local curvature and without quantum fluctuations), 
there is a clear qualitative resemblance with the shapes extracted from the full quantum simulation.
Note that we have not attempted a proper translation between discrete and continuum units of length
and volume, which would be necessary for a quantitative comparison.

%\newpage

\end{document}